\documentclass[12pt]{article}
\usepackage{amsmath, amsthm}
\usepackage{graphicx}
\usepackage{enumerate}
\usepackage{natbib}
\usepackage{url} 

\usepackage{url} 
\usepackage{amsmath}
\usepackage{graphicx}
\usepackage{amsmath, amssymb}
\usepackage{bbold}
\usepackage[colorlinks=true,allcolors=blue]{hyperref}
\usepackage{mathbbol}
\usepackage{float}
\usepackage{xcolor}
\usepackage{bm} 
\usepackage{setspace}
\usepackage{multirow}
\usepackage{lscape}
\usepackage{rotating}
\usepackage{adjustbox}
\usepackage{booktabs}
\usepackage{ragged2e}
\usepackage{float}

\usepackage{amsfonts,amsmath,amssymb}
\usepackage{enumerate}
\usepackage{times}
\usepackage{bm}
\usepackage{natbib}
\usepackage[percent]{overpic}
\usepackage{float}
\usepackage{graphicx}
\usepackage[margin=0pt]{subfig}
\usepackage[plain,noend]{algorithm2e}
\usepackage{lipsum}
\usepackage{bm}
\usepackage{natbib}
\usepackage[plain,noend]{algorithm2e}
\usepackage[utf8]{inputenc}
\usepackage{amssymb}
\usepackage[justification=centering]{caption}
\usepackage{mathrsfs}
\usepackage{graphicx}
\usepackage{mathtools}
\usepackage{booktabs}
\usepackage{amstext} 
\usepackage{epstopdf}
\usepackage{array} 
\usepackage{booktabs}
\usepackage{caption}
\usepackage{epstopdf}
\usepackage{xcolor}
\usepackage{comment}

\newtheorem{theorem}{Theorem}[]
\newtheorem{corollary}{Corollary}[theorem]
\newtheorem{lemma}[theorem]{Lemma}
\theoremstyle{definition}
\newtheorem{definition}{Definition}[section]

\makeatletter
\renewcommand{\algocf@captiontext}[2]{#1\algocf@typo. \AlCapFnt{}#2} 
\def\@algocf@capt@plain{top}
\renewcommand{\algocf@makecaption}[2]{%
  \addtolength{\hsize}{\algomargin}%
  \sbox\@tempboxa{\algocf@captiontext{#1}{#2}}%
  \ifdim\wd\@tempboxa >\hsize
    \hskip .5\algomargin%
    \parbox[t]{\hsize}{\algocf@captiontext{#1}{#2}}
  \else%
    \global\@minipagefalse%
    \hbox to\hsize{\box\@tempboxa}
  \fi%
  \addtolength{\hsize}{-\algomargin}%
}
\makeatother

\def\T{{ \mathrm{\scriptscriptstyle T} }}

\newcommand{\black}[1]{\textcolor{black}{#1}}

\def\independenT#1#2{\mathrel{\rlap{$#1#2$}\mkern2mu{#1#2}}}

\newcommand{\calG}{\mathcal G}
\newcommand{\calI}{\mathcal I}

\newcommand{\calL}{\mathcal L}
\newcommand{\calV}{\mathcal V}

\newcommand{\calD}{\mathcal D}

\newcommand{\bw}{ {\boldsymbol w} }
\newcommand{\bW}{ {\boldsymbol W} }

\newcommand{\given}{\,|\,}

\newcommand{\bphi}{ {\boldsymbol \phi} }

\newcommand{\ind}{\overset{\mbox{ind}} \sim}

\DeclareMathOperator{\tr}{tr}

\begin{document}

\def\spacingset#1{\renewcommand{\baselinestretch}%
{#1}\small\normalsize} \spacingset{1}

 \title{\bf Graph-constrained Analysis for Multivariate Functional Data}
\author{Debangan Dey$^{1}$,
Sudipto Banerjee$^{2}$, 
Martin Lindquist$^{3}$,
and Abhirup Datta $^{3}$\\
$^{1}$ National Institute of Mental Health \\
$^{2}$ Department of Biostatistics, University of California Los Angeles\\
$^{3}$Department of Biostatistics, Johns Hopkins Bloomberg School of Public Health
}

\maketitle





\label{firstpage}

    \begin{abstract}
       \small 
The manuscript considers multivariate functional data analysis with a known graphical model among the functional variables representing their conditional relationships (e.g., brain region-level fMRI data with a prespecified connectivity graph among brain regions). Functional Gaussian graphical models (GGM) used for analyzing multivariate functional data customarily estimate an unknown graphical model, and cannot preserve knowledge of a given graph. We propose a method for multivariate functional analysis that exactly conforms to a given inter-variable graph. We first show the equivalence between partially separable functional GGM and graphical Gaussian processes (GP), proposed recently for constructing optimal multivariate covariance functions that retain a given graphical model. The theoretical connection helps \black{to design} a new algorithm that leverages Dempster's covariance selection for obtaining the maximum likelihood estimate of the covariance function for multivariate functional data under graphical constraints. We also show that the finite term truncation of functional GGM basis expansion used in practice is equivalent to a low-rank graphical GP, which is known to oversmooth marginal distributions. To remedy this, we extend our algorithm to better preserve marginal distributions while respecting the graph and retaining computational scalability. The benefits of the proposed algorithms are illustrated using empirical experiments and a neuroimaging application. 
    \end{abstract}
   
    \noindent%
{\it Keywords:} 
    Gaussian Processes; Conditional independence; Multivariate analysis; Functional data analysis; Graphical models; Spatial data.

\newpage

\section{Introduction}
\label{sec:intro}

Gaussian graphical models (GGM) are extensively used to represent sparsity structures emerging from conditional independence relations among a collection of Gaussian random variables. 
The graph 
specifies the zeroes in the precision 
Conducting multivariate analysis under a known graph constraint, i.e., when the variables conform to a given graph, is of key interest. Covariance selection \citep{dempster1972covariance} offers a seminal result to answer this specific question. 
When the graph is unknown, classical approaches such as the graphical Lasso estimate the graph (from the sparse precision matrix) using L-1 regularization \citep{friedman2008sparse,meinshausen2006high}. 
Bayesian alternatives employ Markov chain Monte Carlo (MCMC) algorithms exploiting a conjugate hyper-inverse Wishart prior for estimating the graph and the precision matrix \citep{dobra2011bayesian, roverato2002hyper, carvalho2007simulation}.

Graphical Gaussian models are customarily applied to multivariate data sets, where the variables represented by the nodes of the graph are scalar- or vector-valued. Here we consider a collection of functions that conceptually exist in a continuum over a domain (e.g., over space or time) and are associated among themselves via a posited conditional independence graph. We consider analysis of partial realizations of these multiple functions or processes, \black{which is always the case in practice.} Examples of such multivariate functional data include physical activity or heart rate continuously measured from actigraph monitors (wearables), EEG/fMRI signals from different brain regions (neuroimaging), multiple-gene expressions from cells distributed over a tissue (spatial transcriptomics), population counts of multiple biological species over a region (ecology) and multiple pollutants measured from environmental monitoring stations (environmental sciences). 

Multivariate functional data are often modeled using a latent multivariate stochastic process, such as a multivariate Gaussian process (GP) (\cite{wackernagel2013multivariate, banerjee2014hierarchical, creswikle11, qiao2019functional}). Hence, it is natural to seek extensions of graphical models to settings where each node represents an entire function or stochastic process. Two recent approaches have extended the GGM to represent conditional dependencies among multivariate stochastic processes on continuous domains: (i) functional GGM (FGGM); and (ii) graphical Gaussian Process (GGP). The former relies on a multivariate Kosambi-Karhounen-Lo\`eve basis expansion and estimates the sparsity (graphical model) of the precision matrices of the Gaussian coefficient vectors via penalized estimation \citep{zapata2019partial, zhu2016bayesian}. Functional GGMs have been primarily used for functional data analysis, where basis function representations \black{are customarily used} but have also been applied to spatial data \citep{krock2021modeling}. These methods estimate the unknown graph through the aforementioned regularization techniques. 

Graphical GP \citep[GGP,][]{dey2021ggp},  originally proposed for multivariate spatial analysis, offers a multivariate covariance function that exactly conforms to a specified graph among the variables. Given this undirected graph and a multivariate covariance function, a GGP is the \black{derived} optimal covariance function that satisfies the prespecified graphical constraint. This means that the processes comprising a GGP exactly satisfy process-level conditional independence specified by the graph. At the same time, GGP retains marginal process distributions from the original multivariate covariance function it is derived from, thereby facilitating the interpretation of the attributes of each individual process via inference on the respective parameters, which is often the primary inferential objective in spatial analysis. GGP is implemented via an algorithm called {\em stitching} that scales effectively for likelihood-based analysis of {\em highly multivariate} spatial data involving a large number of outcome variables by leveraging the sparsity of the graph. 

We focus on multivariate functional data analysis when the inter-variable graph is known 
(e.g. existing brain-region networks, phylogenetic trees etc.), and needs to be preserved in the statistical model. This problem is an extension of Dempster's covariance selection from finite-dimensional covariance matrices to infinite-dimensional covariance operators. In other words, the goal is to find the best estimate of the multivariate covariance function under a graphical constraint on the variables. While having the knowledge of a graph may appear simpler than requiring the estimation of an unknown graph, current FGGM approaches that estimate the graph simultaneously with other parameters cannot preserve a given graphical constraint. Thus, akin to the typical challenges of parameter estimation under a known but complicated constraint, existing FGGM algorithms are not typically conformable to a known graphical constraint. This motivates our current developments. 

Our contributions are summarized as follows. We first draw theoretical connections between functional GGM and graphical GP. These are apparently complementary methods for introducing graphical models in multivariate GP. The former relies on basis function expansions and is typically used for analyzing functional data. Graphical GP focuses on spatial data and directly uses parametric covariance functions. We bridge these two seemingly disconnected paradigms. We formally establish in Theorem~\ref{thm:ggp-fggm-infinite} that the class of {\em partially separable} functional GGM with the graphical model for each of the coefficient covariance matrices is equivalent to a graphical GP. Furthermore, using Nystr\"om's method, we relate the stitching algorithm for constructing a graphical GP to a finite-term truncation of basis functions used in functional GGM. Specifically, we show that truncation corresponds to use of a low-rank GP in stitching that is known to oversmooth. 

The theoretical connections we offer in this manuscript carry important implications for practice in multivariate functional analysis. We first leverage the equivalence between functional GGM and graphical GP to design an algorithm for analyzing multivariate functional data that exactly preserves the given functional conditional independence relationships. In particular, we provide a solution to Dempster's covariance selection for multivariate functional data by deriving the maximum likelihood estimates for a multivariate covariance function under the graphical constraint (Theorem \ref{thm:mle}). We subsequently show that the finite-term truncation of the basis function leads to over-smoothing as it corresponds to stitching using a low-rank GP. We use ideas from constructing full-rank graphical GP via stitching to offer an improved algorithm that mitigates the effects of over-smoothing by better preserving marginal distributions while retaining the graphical constraint. We show the utility of our method through numerous experiments on synthetic data. We demonstrate the application of the method for the analysis of brain region-level functional Magnetic Resonance Imaging (fMRI) data using a connectivity graph among brain regions. \textcolor{black}{The data is from a study of effects of thermal pain on the brain. This application is particularly well-suited for the proposed method as it is known apriori that there are a number of brain regions that are consistently involved with pain processing whose connections together are predictive of physical pain \citep{lindquist2017group,nath2023machine}.}

\section{The two paradigms of graphical functional modeling}\label{sec:rev}
We offer a brief review of functional Gaussian graphical models and graphical Gaussian Processes and how they differ in terms of the model forms and their applications.

\subsection{Functional Gaussian graphical models}\label{sec:fggm}
We first review the definition of conditional independence for \black{Gaussian} processes \black{(GP)} on continuous domains. For a set $\calV=\{1,\ldots,q\}$, let %
$w(\cdot)=(w_1(\cdot),\ldots,w_q(\cdot))^{\T}$ be a $q\times 1$ GP over a continuous domain $\calD$ 
with node $j$ representing the univariate process $w_j(\cdot)$. 
 Two univariate GPs $w_{i}(\cdot)$ and $w_{j}(\cdot)$ are conditionally independent, given the remaining processes $\{w_k(\cdot) \given k \in V \setminus \{i,j\}\}$ if $\mbox{Cov}(\epsilon_{iB}(s), \epsilon_{jB}(s'))=0$ for all $s,s'\in \calD$ and $B = \calV \setminus \{i,j\}$, where $\epsilon_{kB}(s) = w_k(s) - \mbox{E}[w_k(s)\given \sigma(\{w_j(s'): j\in B,\; s'\in \calD\})]$ \citep{dey2021ggp} and $\sigma(\cdot)$ is the $\sigma$-algebra generated by its argument. 
Functional GGM (FGGM) introduces this process-level conditional independence through the precision matrices of coefficient vectors in its basis expansion. Letting $\phi_{jl}$ be the orthonormal basis functions in the domain $\calD$, the process is represented 
as $w_j(s) = \sum_{l=1}^{\infty} \theta_{jl} \phi_{jl}(s)$. 
\cite{zhu2016bayesian} show that modeling the coefficients $\theta_{jl}$ to be Gaussian and assuming a sparsity structure between the discrete multivariate GP $\theta^{(j)}=\{\theta_{jl}: l=1,\cdots, \infty\}$ imposes the corresponding process-level conditional independence among the component processes of $w(\cdot)$. In practice, they truncate the expansion of \black{$w_{(j)}(\cdot)$} to $m_{j}$ terms and fit a GGM on the $M=\sum_{j=1}^{q} m_{j}$ dimensional vector of coefficients.
The approach involves inverting an $M \times M$ matrix, thereby requiring $O(M^3)$ floating point operations (FLOPs) or $O(q^3 m^3)$ FLOPs when $m_j = m$ for all $j$. The computational expense being cubic in the number of basis functions $m$ prohibits large values for $m$, which, in turn, restricts the richness of the function class.  

The functional GGM in \cite{zapata2019partial} and \cite{krock2021modeling} introduces two further assumptions: (a) a common orthonormal basis function, i.e., $\phi_{jl} = \phi_{l}$ for all $j=1, \cdots, q$; and (b) the coefficients corresponding to different basis functions are independent, i.e, ${\theta}_j$ and ${\theta}_k$ are independent, where ${\theta}_j= (\theta_{1j}, \cdots, \theta_{qj})$.  Due to the separability in the joint covariance structure of basis coefficients, they are formally termed as {\em partial separable} processes. The basis function expansion is truncated to $m$ terms in practical implementation. This partial separable model only needs $m$ inversions of $q \times q$ matrices ($Cov(\theta_j), j=1,\ldots,m$)  and considerably reduces computational complexity to $O(mq^3)$, thereby allowing $m$ to be larger. The algorithm maximizes a penalized likelihood with a graphical Lasso penalty on the estimated covariance of the coefficient vectors to induce sparsity. This results in the estimation of an unknown graphical model corresponding to the common sparsity among the precision matrices for each of the coefficients. Recently, \cite{li2018nonparametric, lee2022nonparametric} proposed a class of non-parametric graphical models for functional data that performs better than FGGMs when the relation between variables is non-linear or heteroskedastic in nature.

\subsection{Graphical Gaussian  Processes}\label{sec:ggp}

Spatial data 
is  
often modeled using GPs with parametric covariance functions. 
The Mat\'ern family of covariances \citep{stein1999interpolation} are popular in spatial analysis because their parameters are interpreted as the variance, spatial smoothness, and range of the process. For multiple dependent outcomes $w(\cdot)=(w_1(\cdot),\ldots,w_q(\cdot))^{\T}$, a common GP specification is $w(\cdot) \sim GP(0,C(\cdot,\cdot))$ where $C=(C_{ij})$ is the $q \times q$ multivariate covariance function. The multivariate Mat\'ern is a popular choice for $C$, ensuring that the derived covariance and cross-covariance functions are also Mat\'ern \citep{gneiting2010matern,apanasovich2012valid}, allowing us to interpret each component process. However, the model struggles to scale up to highly multivariate (large $q$) settings.

\cite{dey2021ggp} introduced graphical Gaussian process (GGP) models for analyzing highly multivariate spatial data. These are specified through multivariate covariance functions that exactly encode an inter-variable graphical model. Theorem~1 in \cite{dey2021ggp} asserts that given any non-graphical stationary multivariate GP and any inter-variable graph, there is a unique and optimal GP approximating the original GP (in terms of integrated spectral Kullback-Leibler divergence) while respecting process-level conditional independence relationships as specified by the graph.  This optimal {\em graphical GP} was shown to retain the univariate marginal distributions from the original GP exactly. For any pair of variables $(i,j)$ included in the edge set of the graph, the cross-covariance is also retained from the original GP. 
We denote this optimal graphical GP derived from the original covariance function $C$ and a graph $\calG$ as $GGP(C,\calG)$.

\begin{definition}[Graphical GP]\label{def:ggp} $w(\cdot) \sim GGP(C,\calG)$ is a $q \times 1$ graphical GP on $\calD$ given a valid $q \times q$ cross-covariance function $C=(C_{ij})$ on $\calD \subseteq \mathbb R^d$ and a graph $\calG=(\calV,E)$, such that 
\begin{enumerate}
    \item Retains univariate marginal distributions: $w_j(\cdot) \sim GP(0,C_{jj})$ for all $1 \leq j \leq q$,
    \item Retains edge-specific cross-covariances: $Cov(w_i(s),w_j(t))=C_{ij}(s,t)$ for all $(i,j) \in E$ and $s,t \in \calD$,
    and 
    \item Encodes process-level graphical model as in Definition \ref{def:ggp}: $w_i(\cdot) \perp w_j(\cdot) \mid w_{-ij}(\cdot)$ for all $(i,j) \notin E$.
\end{enumerate}   
\end{definition}

\black{Construction of the graphical GP used an algorithm called ``stitching''. The stitching algorithm in \cite{dey2021ggp} combines two algorithms, which we refer to as ``stitching" and ``stretching" in the current manuscript.} \black{The step of stitching} is applying Dempster's covariance selection using $\calG$ on the covariance matrix arising from $C$ when considering a finite set of locations \black{$\calL = \{s_1, \cdots, s_p\}$. This defines a graphical GP on the finite set $\calL$ that adheres to the conditional independences specified via the graphical model $\calG$. The stretching step extends the finite-dimensional process to the continuous infinite domain by adding a residual error process, while preserving the graphical model. }

\subsubsection{Stitching} 

\black{To perform the stitching}, we need to define covariance selection for positive definite (p.d.) matrices.
\begin{definition}[Covariance Selection,  \cite{dempster1972covariance, speed1986gaussian}]\label{def:covsel}
Given a graph $\calG=(\calV,E)$ and a \black{positive definite} matrix $A$ with row and column blocks indexed by $\calV$, $CovSel(A,\calG)$ is the unique p.d. matrix $B$ from Dempster's covariance selection on $A$ using $\calG$, i.e., $B_{ij}=A_{ij}$ for $i=j$ or $(i,j) \in E$, and $(B^{-1})_{ij}=O$ for $(i,j) \notin E$.
\end{definition}
 Given $C$, $\calG$, and $\calL$, stitching first defines a graphical Gaussian model (GGM) on $\calL \subset \calD$.  
\begin{equation}\label{eq:covsel}
    w_{stitch}(\calL) \sim N(0,\tilde  C(\calL,\calL)) \mbox{ where } \tilde  C(\calL,\calL)= Covsel(C(\calL,\calL),\calG).
\end{equation}
\black{Here, $C(\calL, \calL)$ denotes the covariance matrix obtained from evaluating the covariance function $C$ on the reference set $\calL$.} 

\subsubsection{\black{Stretching}}\label{sec:stretch}

\black{As part of the stretching,} the covariance matrix on a finite set is then extended to a covariance function on the infinite domain. In other words, the finite-dimensional GGM (\ref{eq:covsel}) is extended in \cite{dey2021ggp} to an infinite-dimensional GGP on $\calD$ as 
\black{
 \begin{equation}\label{eq:stitch} 
 \hspace{-4.5mm} w_{stretch,j}(s) = w^{pp}_{stitch,j}(s) + z_j(s), \mbox{ where } w^{pp}_{stitch,j}(s)=C_{jj}(s,\calL)C_{jj}(\calL,\calL)^{-1}w_{stitch,j}(\calL).
 \end{equation} 
 }
 Here, \black{$C_{jj}(s, \calL) = (C_{jj}(s,s_1) , \cdots, C_{jj}(s,s_p))$ } and $w^{pp}_{stitch}$ is the (fixed-) or low-rank {\em predictive process} based on the finite-dimensional distribution $w_{stitch}(\calL)$ defined on the set of {\em knots} $\calL$ \citep{banerjee2008gaussian}. The processes $z_j(\cdot) \overset{\mbox{ind}_j}{\sim} GP(0,C_{jj,res})$ are the residual GP with $C_{jj,res}(s,t) = C_{jj}(s,t) - C_{jj}(s,\calL)C_{jj}(\calL,\calL)^{-1}C_{jj}(\calL,t)$. Decomposing a univariate GP into a predictive process and an orthogonal residual process is well known \citep{banerjee2008gaussian}. The {stitching and stretching process} introduced two innovations to this decomposition. First, the process on $\calL$ is endowed with the covariance $\tilde C(\calL,\calL)$ instead of $C(\calL,\calL)$. This ensures that the graph $\calG$ is encoded while preserving marginals on $\calL$. Orthogonality of $z_j(\cdot)$ and $w^{pp}_{stitch,j}(\cdot)$ then implies that the marginals are preserved over entire $\calD$ if the residual processes, $z_j(\cdot)$'s, are added to the low-rank Gaussian process $w^{pp}_{stitch,j}(s)$, \black{i.e., $w_{stretch,j}(\cdot) \sim GP(0,C_{jj})$ for all $j$ on the entire domain $\calD$.} Second, the $z_j(\cdot)$'s are chosen to be independent component-wise (across $j$'s), ensuring that the GGM encoded over $\calL$ is extended to process-level conditional independence relations over $\calD$ yielding the GGP, \black{i.e., for all $(i,j) \notin E$, we have $w_{stretch,i}(\cdot) \perp w_{stretch,j}(\cdot) \given w_{stretch,-ij}(\cdot)$.} For a decomposable sparse graph, a GGP likelihood from this process involves considerably less parameters and FLOPs than a full GP with covariance $C$. 

\section{Theoretical connections between functional GGM and graphical GP}\label{sec:theory}

The GGP in \cite{dey2021ggp} is constructed from a parametric multivariate covariance function $C$ such as the multivariate Mat\'ern. In analyzing functional data, covariance functions are often represented more generally using basis functions. 
Here, we demonstrate how a graphical GP can be derived from \black{any multivariate GP that admits a} partially separable basis function expansion. The resulting GGP \black{is a FGGM that} will exactly conform to a given graph, and we show, in Section \ref{sec:methods}, how this is leveraged to conduct graph-constrained analysis of multivariate functional data. 

\subsection{Covariance selection on partially separable Gaussian processes}

Consider a $q \times 1$ partially-separable multivariate GP on a continuous domain $\calD$,
\begin{equation}\label{eq:ps}
w_0(s)= \sum_{l=1}^{\infty} {\theta}_l \phi_l(s),\mbox{ where } {\theta}_l = (\theta_{1l}, \cdots, \theta_{ql}) \sim N_q (0,\Sigma_l).
\end{equation}

Theorem~$\ref{thm:ggp-fggm-infinite}$  proves that the graphical GP derived from $w_0(\cdot)$ is a partially separable functional GGM that exactly preserves a given graphical model.

\begin{theorem}\label{thm:ggp-fggm-infinite}
Let $C$ be the cross-covariance function of a stationary partially separable process $w_0(\cdot)$ in (\ref{eq:ps}). Consider a new partially separable process  
$w(s) = \sum_{l=1}^{\infty} {\theta}_l^* \phi_l(s)$ where ${\theta}_l^* = (\theta_{1l}^*, \cdots, \theta_{ql}^*) \sim N_q (0,\Sigma_l^*)$. Then for any fixed graph $\calG = (\calV, E)$,  $w(\cdot) \sim GGP(C,\calG)$ as in Definition \ref{def:ggp} if and only if $ \Sigma^*_l=\mbox{CovSel}(\Sigma_l,\calG)$ for all $l$.
\end{theorem}

The result shows that starting with a partially separable GP, the optimal graphical GP can be obtained by applying Dempster's covariance selection on each covariance matrix  $\Sigma_l$ 
of the basis coefficients, and this graphical GP is a partially separable functional GGM. All proofs are provided in the Appendix. 

\subsection{Stitching and truncation}\label{sec:stitchtrunc}
In practice, we cannot work with infinite basis expansions \black{and construct a process using only a finite set of orthogonal functions extracted from} 
the graphical GP $w(\cdot)$ in Theorem \ref{thm:ggp-fggm-infinite}. 
The following result is an immediate consequence of Theorem~\ref{thm:ggp-fggm-infinite}. \black{The proof is outlined in Section \ref{sec:cor-proof}}. 
\begin{corollary}\label{cor:trunc}
Let $w_{0}^m(s)$ be constructed using $m$ basis functions of a graphical GP as
\begin{equation}\label{eq:trunc}
w_{0}^m(s)=\sum_{l=1}^m \theta_l \phi_l(s), \black{\mbox{ where } {\theta}_l = (\theta_{1l}, \cdots, \theta_{ql}) \sim N_q (0,\Sigma_l)}\;  
\end{equation}
and $C^m$ is the covariance function of $w_{0}^m(s)$. Then $w^{m}(\cdot) \sim GGP(C^m,\calG)$ is given by $w^{m}(s)= \sum_{l=1}^{m} {\theta}_l^* \phi_l(s)$  where  $Cov(\theta^*_l)=Covsel(\Sigma_l, \calG)$ for  $l=1,\ldots,m$. 
\end{corollary}

We discuss in Section~\ref{sec:methods} how this connection facilitates a practical estimation strategy for finite-term functional GGM that exactly conforms to a given graph. However, a finite-term truncation can come at the expense of reduced accuracy in estimating the covariance function. To explicitly demonstrate the impact of truncation, we now relate this truncated functional GGM $w^{m}(\cdot)$ of Corollary \ref{cor:trunc} to a low-rank stitched graphical GP (recall Section~\ref{sec:ggp}) through an approximate result using N\"ystrom approximation for kernel matrices \citep{drineas2005nystrom}. 
Let $w_0(\cdot)$ 
be the partially separable GP in (\ref{eq:ps}) with cross-covariance function $C=(C_{ij})$, $w(\cdot)$ be the partially separable $GGP(C,\calG)$ derived from $w_0(\cdot)$ using Theorem~\ref{thm:ggp-fggm-infinite} and $w^m(\cdot) $ be the finite-term truncation of $w(\cdot)$ in Corollary \ref{cor:trunc}. We show in Section \ref{sec:lowrank} of the Appendix that $w^m(\cdot)$ is approximately equal to $w^{pp}_{stitch}(\cdot)$, where $w^{pp}_{stitch}=(w^{pp}_{stitch,1}(\cdot),\ldots,w^{pp}_{stitch,q}(\cdot))$ is the low-rank predictive process \black{defined} in (\ref{eq:stitch}) created as part of the stitching \black{and stretching} constructions.

\black{Theorem 1 proves the equivalence between the theoretical constructions of functional GGM and the graphical GP. Similarly, the above approximate result establishes an analogy between the practical procedures of truncation and stitching for the two approaches.}
\black{This result also highlights that truncation ignores the equivalence of the stretching step employed earlier in graphical GP (Section \ref{sec:stretch}). The stitched and stretched graphical GP $w_{stretch}(s)$}, as specified in (\ref{eq:stitch}), is the sum of the stitched low-rank predictive process $w^{pp}_{stitch}(s)$ and the independent residual processes $z(\cdot)$, i.e.,  \black{$w_{stretch}(s) = w^{pp}_{stitch}(s) + z(s)$}, where $z(s)=(z_1(s),\ldots,z_q(s))$. The truncated functional GGM $w^m(s)$ is only equivalent to the first component $w^{pp}_{stitch}(s)$ and ignores the residual component $z(s)$. This difference has important \black{practical} implications.  The marginal distributions of the process $w^{pp}_{stitch}(\cdot)$ has the same distribution as  $C_{jj}(\cdot,\calL)C_{jj}(\calL,\calL)^{-1}w_{0,j}(\calL)$ --- the univariate predictive processes of \cite{banerjee2008gaussian} derived from the original process $w_0(s)$. These low-rank predictive processes have been shown, theoretically and empirically, to underestimate the marginal variances of the original process $w_0(\cdot)$, resulting in oversmoothing and inferior performance  \citep{stein2014limitations, datta2016hierarchical}. 
As $w^m(\cdot)$ 
 is equivalent to $w^{pp}_{stitch}(\cdot)$, the truncated process $w^m(\cdot)$ \black{also will} not preserve the marginals of the original distribution $w_0(\cdot)$. Rather $w^m(\cdot)$ has the same marginal distributions as the predictive process, \black{and will be expected to oversmooth the marginal spatial patterns. In contrast, the process $w_{stretch}(\cdot)$}, on account of adding the residual processes $z(\cdot)$, preserves the marginals of the original process $w_0(\cdot)$ and would not oversmooth. Thus, we see a potential drawback of the truncation. We demonstrate the oversmoothing in our empirical results and in Section \ref{sec:methods} we offer a solution to the issue.

\section{Estimation}\label{sec:methods}
In scientific applications, we either know the graph apriori or have to estimate it. Functional GGM is widely deployed in the \black{latter} setting  \citep{zapata2019partial,krock2021modeling}. However, if the graph is known (e.g., a known phylogenetic tree as a graph to model multiple species distributions over a region, a known auto-regressive time-dependence structure to model spatiotemporal pollutant data, connectivity graph between brain regions estimated from a secondary dataset), current functional GGM approaches cannot explicitly encode this information and preserve the graphical model in the analysis. 

In this Section, we leverage the theoretical connections developed in Section \ref{sec:theory} to present algorithms for graph-constrained analysis of multivariate functional data. 

\subsection{Graph-constrained MLE for multivariate functional data}
Theorem~\ref{thm:ggp-fggm-infinite} and Corollary~\ref{cor:trunc} establish that the optimal approximation of a partially separable process under a graphical constraint is also a partially separable process with Dempster's covariance selection applied to covariance matrices for each of the basis function coefficients. We now provide the data-driven analog of the result that is practically applicable when analyzing multivariate functional data occurring as partial realizations of these multivariate processes. In particular, the following theorem provides the maximum likelihood estimate (MLE) for a partially separable multivariate covariance function under a graphical constraint. The result is viewed as an extension of Dempster's covariance selection, i.e., graph-constrained MLE of covariances, from the setting of vector-valued data with covariance matrices to functional data with covariance operators. 

\begin{theorem}\label{thm:mle}
Let $X_1, X_2, \cdots, X_N$ denote iid realizations of $q$-variate functional data on $p$ locations $\{s_1, s_2, \cdots, s_p\}$, generated from a mean-zero \black{truncated} partial separable \black{GP$(0,C^m)$ as in Corollary (\ref{cor:trunc})} with known orthonormal basis functions $\phi_l$. Then, for a given functional graphical model $\calG$, the graph-constrained MLE of the covariance function \black{$C^m$} is given by 
\begin{equation}\label{eq:mlegraph}
\black{\widehat{C^m}} = \sum_{l=1}^{m} \mbox{CovSel}\big((I_q \otimes \bphi_l') S (I_q \otimes \bphi_l), \calG \big) \otimes \bphi_l  \bphi_l'
\end{equation}
where $S = \frac 1N \sum_{i=1}^N X_i X_i'$ is the sample covariance matrix,  $\boldsymbol{\phi}_l' = (\phi_l(s_1), \cdots, \phi_l(s_p))$ is the basis function  evaluated at $p$ locations, and $CovSel$ is covariance selection (see Definition \ref{def:covsel}). 

\end{theorem}

\begin{lemma}\label{lem:con}
\black{Under the setting of Theorem \ref{thm:mle}, the graph-constrained MLE $\widehat C^m$ in \eqref{eq:mlegraph} is an asymptotically consistent estimator of the covariance operator $C^m$}. 
\end{lemma}

Theorem \ref{thm:mle} provides a direct approach to obtaining the MLE of a multivariate covariance function under a given graphical constraint and \black{Lemma \ref{lem:con} shows the consistency of the estimator. All the proofs are shown in Appendix}. In practice, the basis functions are often chosen in a data-driven manner instead of being fixed apriori. These estimates $\hat\phi_l$ can be obtained from a functional principal components analysis. Plugging these (or any other estimate of $\phi_l$), we can use Theorem \ref{thm:mle} to calculate the profile maximum likelihood estimate (PMLE) of \black{$\Sigma_l$} as $\hat{S}_l= (I_q \otimes \hat{\bphi}_l') S (I_q \otimes \hat{\bphi}_l)$. 
Subsequently the graph-constrained PMLE for \black{$C^m$} 
can be obtained as 
\begin{equation}
\label{eq:estimate-C}
\black{\widetilde{C^m}} = \sum_{l=1}^{m} \mbox{CovSel}(\hat{S}_l, \calG)  \otimes \hat{\bphi}_l \hat{\bphi}_l'.
\end{equation}

The number of basis functions $m$ is usually chosen so that the first $m$ eigenfunctions $\hat \phi_l$'s explain a given percentage \black{$v$} of the total variance. 
We formally lay out the above steps in Algorithm \ref{alg:est-known-graph} below and term our approach as FGGM-CovSel. 

\begin{algorithm}[!h]
\vspace*{-6pt}
\caption{(FGGM-CovSel: Graph-constrained multivariate functional data analysis)}\label{alg:est-known-graph}
\vspace*{-5pt}
\BlankLine
1. Fix a proportion of variance explained threshold, $v$, say, $v=0.95$ \;
2. Perform pooled functional principal component analysis (\cite{yao2005functional, yang2011functional})
to get $m$ common eigenfunction estimates $\hat{\bphi}_l, l=1, \cdots, m$ that explain $v$ proportion of total variance. \; 
3. Use the eigenfunctions $\hat \phi_l$ to get first $m$ functional principal component scores $\hat{{\theta}}_{il}= (I_q \otimes \hat \bphi_l') X_i$ and calculate the corresponding covariance matrix $\hat S_l = \textrm{cov}_N(\hat{{\theta}}_{il}) = (I_q \otimes \hat{\bphi}_l') S (I_q \otimes \hat{\bphi}_l)$ for $l= 1, \cdots, m$ where $S = \frac 1N X_i X_i'$.\;
 4. Using Iterative Proportional Scaling  (IPS) algorithm \citep{speed1986gaussian} for covariance selection to obtain  $\mbox{CovSel}(\hat S_l,\calG)$ for $l=1, \cdots, m$.\; 
 5. Produce the estimate of the covariance operator $\black{\widetilde{C^m}}$ from \eqref{eq:estimate-C}.
\end{algorithm}

\subsection{Improved estimation of marginal distributions using \black{stretching}}

In Algorithm~\ref{alg:est-known-graph}, using a larger \black{$v$} improves the retention of marginal covariance functions but requires a larger number of basis functions $m$. This is prohibitive computationally, as the iterative proportional scaling (IPS) algorithm \citep{speed1986gaussian} for covariance selection on the $S_l$'s has a total computational complexity of $O(mq^3)$ floating point operations or FLOPs for the $m$ terms. 
For large $q$, this mandates choosing a smaller $m$ to reduce computations. This tradeoff leads to a worse approximation of the marginal covariances and oversmoothing as the finite-term truncated process is equivalent to a low-rank \black{GP} approximation at $m$ locations (see the discussion in Section \ref{sec:stitchtrunc}). 

We now offer an extension of Algorithm \ref{alg:est-known-graph} using the principles of \black{stretching} that remains computationally feasible while offering significantly better retention of the marginal distributions and, thereby, less oversmoothing. 
In Section \ref{sec:stitchtrunc}, we have established that 
$\black{w_{stretch}}(\cdot) \approx w^{m}(\cdot) + z(\cdot)$. The graphical GP $\black{w_{stretch}}(\cdot)$ exactly preserves the marginals of the original process $w_0(\cdot)$, while the truncated process $w^{m}(\cdot)$ is equivalent to a low-rank predictive process and can oversmooth. For a univariate predictive process, the oversmoothing is mitigated by adding the residual process \citep{finley2009improving}. \black{Stretching} proceeds similarly in a multivariate graphical model by introducing the residual process $z(\cdot)$ (defined in \eqref{eq:stitch}) to exactly preserve the marginals while still conforming to the graphical constraint. 

We leverage this connection between truncation and stitching  
to remedy the oversmoothing by the process $w^{m}(\cdot)$ truncated at some computationally feasible value of $m$. We modify Algorithm \ref{alg:est-known-graph} to emulate the full rank stitching \black{and stretching} procedure of \cite{dey2021ggp} by adding component-specific residual processes. This essentially adds a block-diagonal function to the estimate of the covariance function that mitigates the bias in the marginal distributions due to truncation. 
To achieve this, we first run FGGM-CovSel for a small or moderate $m$ as afforded by the computing resources.  
Then, we calculate the residuals and perform functional principal component analysis of the component-specific residuals, and add the diagonal matrix of the estimated residual variances to our covariance estimate \black{$\widehat{C^m}$} from Algorithm \ref{alg:est-known-graph}.
This yields significant improvements in retaining the univariate distributions.  The Algorithm is referred to as \black{FGGM-Stretch} and is formally laid out in Algorithm \ref{alg:est-stitch} below. 

\begin{algorithm}[!h]
\caption{(\black{FGGM-Stretch})}\label{alg:est-stitch}
\vspace*{-5pt}
1. Fix a smaller proportion of variance explained threshold, $v$, say, $v=0.75$.\;
2. Run Algorithm $1$ to get estimates $\hat{{\theta}}_{il}, \hat{\bphi}_l, l=1, \cdots, m$ and $\hat{C}$.\;
3. Calculate residuals: $Z_{i} = X_{i} - \sum_{l=1}^{m} \hat{{\theta}}_{il} \hat{\bphi}_l$ for $i=1, \cdots N$\;
4. Perform functional principal component analysis of the residuals of $Z^{(j)} = (Z_{1j}, \cdots, Z_{Nj})$ for $j=1, \cdots, q$ independently for a larger proportion of variance explained, $v'$, say, $v'=0.95$\;
5. Obtain the eigenvalues as $\hat{\lambda}_{lj}, l=1, \cdots, m', j =1, \cdots, q$\;
6. Find the new stitched covariance operator $\black{\widehat{C^m}^{(2)}} = \black{\widehat{C^m}} + \sum_{l=1}^{m'} \textrm{diag}(\hat{\lambda}_{l1}, \cdots, \hat{\lambda}_{lq})$. 
\end{algorithm}

\section{Empirical results}\label{sec:sim}
We perform a series of simulation experiments to gauge the performance accuracy of three algorithms for graph-constrained analysis of multivariate functional data --- functional Gaussian graphical model (FGGM) of \cite{zapata2019partial}, and our two proposed algorithms FGGM-CovSel and \black{FGGM-Stretch}. The first one cannot use the knowledge of the graph and estimate the graph, treating it as unknown while simultaneously estimating the covariance function. The latter two exactly preserve the graph in the analysis. 

\begin{figure}[h]
\centering
\includegraphics[scale=1.2, trim={0 0 0 0},clip]{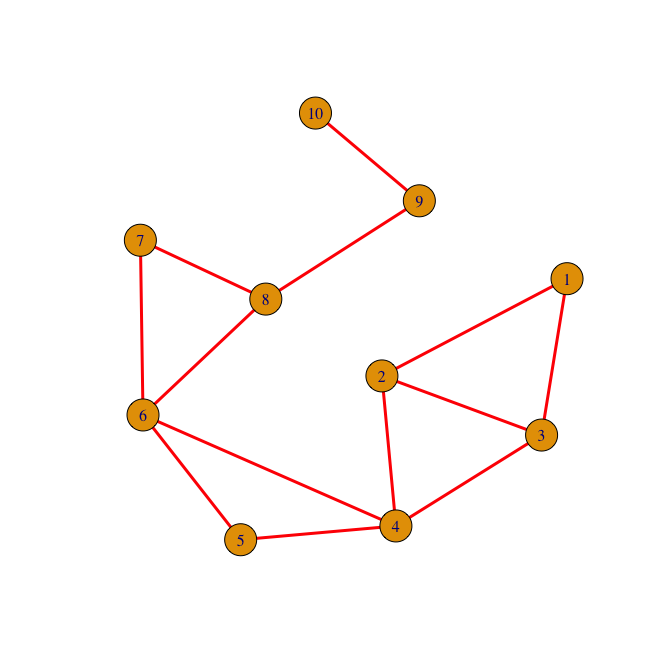}
    \caption{The $10$-variable graph used in the simulation studies.}\label{fig:decomp-g}
\end{figure}

For the data generation, we consider $q=10$-variate GPs adhering to the conditional independence structure from the graph in Figure \ref{fig:decomp-g}. We consider two choices of covariance functions generating the data -- (a) correctly specified case: partial-separable (PS) covariance, and (b)  misspecified case: graphical Mate\'rn (GM) covariance of \citep{dey2021ggp} that is not partially separable. We fix the total number of basis functions at $L=101$ for data generation. We take them to be the Fourier basis functions and evaluate them at $200$ equally spaced points between $(0,1)$. Now to generate the basis-specific coefficients, we generate the covariance matrices as $\Sigma_l=a_l \Omega_l^{-1}$ for $l=1, \cdots, 101$, where the precision matrices $\Omega_l$ are generated using the algorithm of \cite{peng2009partial} to have the exact sparsity as specified by the graph $\calG$ (Figure \ref{fig:decomp-g}). The decaying constants $a_l = 3l^{-1.8}$ ensure that $\tr(\Sigma_l)$ is monotonically decreasing in $l$. The covariance function for the misspecified case was obtained by stitching using multivariate Mat\'ern model \citep{apanasovich2012valid} and the graph $\calG$ (Figure \ref{fig:decomp-g}). For every $(i,j)$ pair of variables, the isotropic multivariate Mat\'ern cross-covariance function on a $d$-dimensional domain is defined as $C_{ij}(s,s')=\sigma_{ij}H_{ij}(\|s - s' \|)$, where $H_{ij}(\cdot)=H(\cdot \given \nu_{ij}, \phi_{ij})$, $H$ being the Mat\'ern correlation function \citep{apanasovich2012valid}. $\sigma_{ij}, \phi_{ij}, \nu_{ij}$ are interpreted as the cross-covariance, cross-scale, and cross-smoothness parameters. To ensure positive-definiteness of cross-covariance matrix, we parametrize as $\sigma_{ij} = b_{ij} \frac{\Gamma(\frac 12 (\nu_{ii}+\nu_{jj} + d))\Gamma(\nu_{ij})}{\phi_{ij}^{2\Delta_A+\nu_{ii}+\nu_{jj}}\Gamma(\nu_{ij} + \frac d2)},$, where $b_{ij}= (\sigma_{ii}\sigma_{jj})^\frac{1}{2} \phi_{ii}^{\nu_{ii}} \phi_{jj}^{\nu_{jj}} \frac{r_{ij}}{\Gamma(\nu_{ij})}$, and $R = (r_{ij}) > 0$, is correlation matrix. For this simulation, we take $\nu_{ij}=\nu_{ii}=\nu_{jj} = \frac{1}{2}$, and $\Delta_A=0$  and $\phi^2_{ij}=(\phi^2_{ii}+\phi^2_{jj})/2$. The marginal range parameters $\phi_{ii}$ and variance parameters $\sigma_{ii}$ were permutations of equispaced numbers in $(1,5)$, while the $R=r_{ij}$'s were chosen as a random correlation matrix. For data generation locations, we take $250$ points on the real line equispaced between $(0,1)$. 

We compare the estimated marginal and edge-specific cross-covariance surfaces qualitatively using some heat-maps and quantitatively using the distance between true and estimated covariance sub-matrices in terms of Kullback-Leibler (KL) divergence. The KL divergence between two matrices $A$ and $B$ is defined as- $\textrm{KL}(B\mid\mid A) = \frac{1}{2}(\textrm{tr}(A^{-1}B) - log(\frac{\mid B\mid}{\mid A \mid}))$. In case (a), which is the truly specified case, we present the results comparing FGGM-CovSel and \black{FGGM-Stretch} with the FGGM algorithm of \cite{zapata2019partial} (implemented using the R package \emph{FGM} \citep{fgm}). \black{We report the graph estimated through FGGM and how it compares with the true graph in Figure \ref{fig:graph-est} in the Appendix. The figure shows that FGGM estimates the graph better in a correctly specified setting (set (a)) than in a misspecified setting (set (b)).}

\begin{figure}[h]
    \begin{center}
\includegraphics[scale=0.8,trim={0 0 0 0cm},clip]{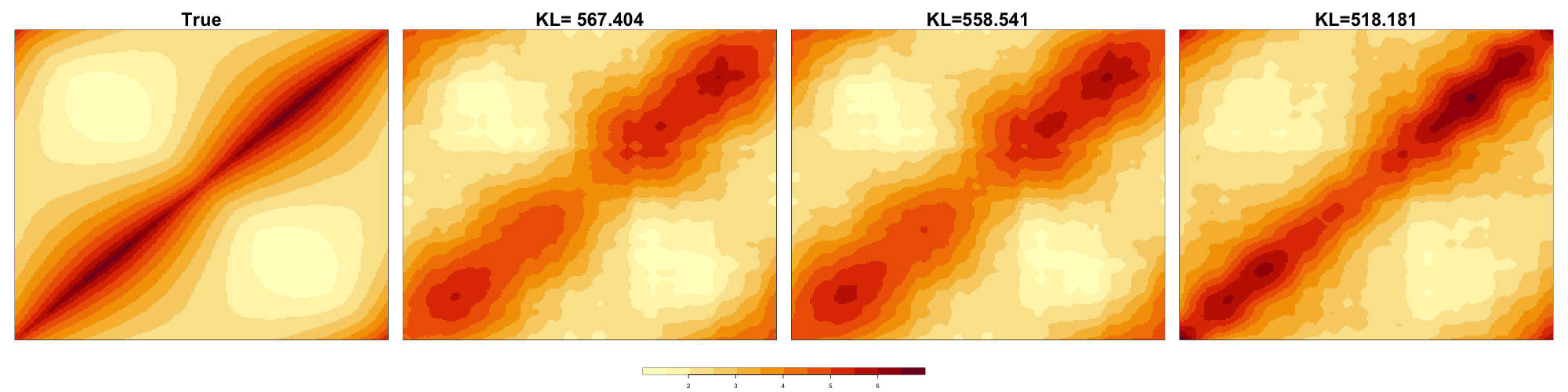}
\end{center}
 \caption{\black{Marginal covariance surfaces for variable $1$: True, FGGM, FGGM-CovSel, \black{FGGM-Stretch} (from left to right). The KL-divergence for each method is noted at the top of the respective panel.}}
    \label{fig:sim-res-stitch}
\end{figure}

 We first compare the true and estimated marginal covariance functions under the correctly specified setting. In Figure \ref{fig:sim-res-stitch}, we plot the true marginal covariance for variable $1$ and the estimated covariance from each of the three methods. We see that \black{FGGM-Stretch} has a significantly improved estimate of the marginal covariance compared to both FGGM and FGGM-CovSel. The FGGM and FGGM-CovSel rely on a low-rank truncation, which evidently leads to oversmoothing, while \black{FGGM-Stretch} mitigates this issue to a great extent. This aligns with our theoretical insights from Section \ref{sec:stitchtrunc}. The result holds for the marginal covariance of all the variables as demonstrated in Figure \ref{fig:sim-res-stitch-rest} in Appendix. 

\begin{figure}[h]
    \begin{center}
        \subfloat[\black{Set (a): The estimated cross-covariance surface of $(1,2)$ edge: (from left to right) True, FGGM, FGGM-CovSel (which has same cross-covariances as \black{FGGM-Stretch})} ]{\includegraphics[scale=0.8,trim={0 0 0 0cm},clip]{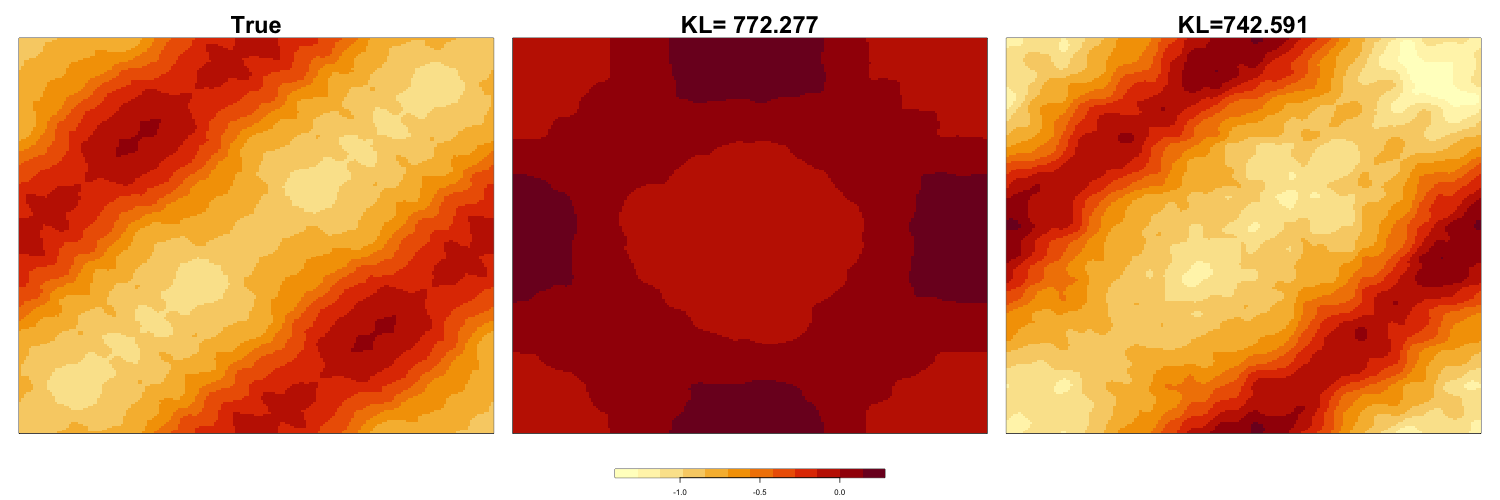}\label{fig:sim-ps}}\hspace{0cm}
        \subfloat[\black{Set (b): The estimated cross-covariance surface of $(1,2)$ edge: (from left to right) True, FGGM, FGGM-CovSel (which has same cross-covariances as \black{FGGM-Stretch}) }]{\includegraphics[scale=0.8,trim={0 0 0 0cm},clip]{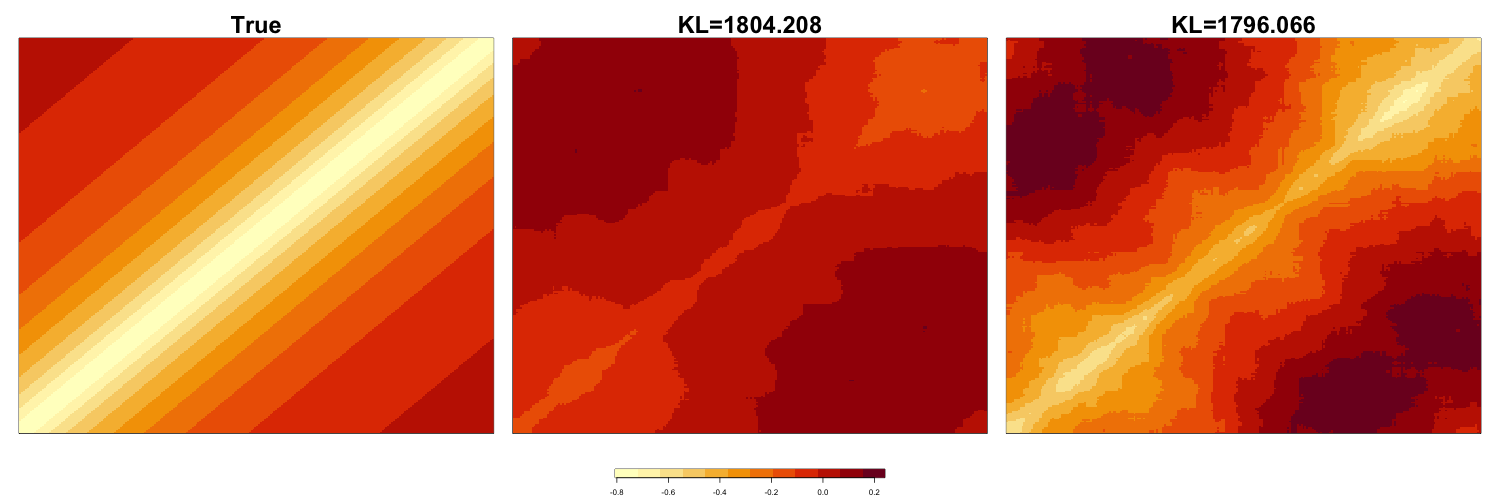}\label{fig:sim-gm}}\hspace{0cm}
        \end{center}
    \setlength{\belowcaptionskip}{-10pt}
\setlength{\abovecaptionskip}{-0pt}    
    \caption{\black{Comparison of $(1,2)$ variable pair cross-covariance surface heatmaps under set (a) and set (b). The KL-divergence for each method is noted at the top of the respective panel.}}
    \label{fig:sim-res}
\end{figure}

We then look at the estimates of the cross-covariance surfaces in Figure \ref{fig:sim-res}. These will be the same for the proposed algorithms FGGM-CovSel and \black{FGGM-Stretch}, which differ only on the marginal covariances. We see that, for the correctly specified case, this estimate from our proposed algorithms performs better than the FGGM estimate (Figure \ref{fig:sim-ps}) and estimates the true cross-covariance surface with high accuracy. This behavior is expected since the model is correctly specified for our algorithms, and the FGGM algorithms do not utilize knowledge of the graph but estimate it. Figure \ref{fig:sim-gm} presents the cross-covariances for the mis-specified scenario. 
We see that FGGM-CovSel estimates the true covariance function less accurately than in the correctly specified case but does capture the broad features like the banding (decay of the Mate\'rn cross-covariance away from the diagonal). This decrease in estimation accuracy is expected since we're trying to fit a misspecified partial separable cross-covariance to a non-partially separable GP. However, even in the misspecified case, FGGM-CovSel still significantly outperforms FGGM. 

Overall, across all edges and in both the correctly specified and misspecified scenarios \black{(across 25 different seeds)}, FGGM-CovSel reports uniformly lower KL divergence between true and estimated covariance matrices than FGGM (Table \ref{tab:kl}). \black{Each row in Table \ref{tab:kl} corresponds to the KL divergence mean (sd) across 25 replicates between edge-specific true and estimated covariance sub-matrices. We observe that} FGGM-CovSel performs according to our expectations in the \black{correctly specified example}, thus validating the theoretical basis of our approach. \black{Moreover, in the correctly specified setting, we perform additional simulation experiments with increasing sample size ($N$ varies from $50$ to $5000$) to demonstrate the asymptotical consistency of FGGM-CovSel estimator. In Figure \ref{fig:asymp-frob} in Appendix, we see decrease in the Frobenius norm between true and estimated covariance with both increasing number of basis functions and number of samples (N)}. \black{The better performance of FGGM-CovSel in the misspecified example (set (b)) gives us confidence in our approach even when the partial separability assumption does not hold}. This is in addition to the benefit demonstrated in Figure \ref{fig:sim-res-stitch} of using \black{FGGM-Stretch} to better capture the marginal distributions.

\begin{table}[ht]
\centering
\color{blue}{\begin{tabular}{|rl|rr|rr|}
  \hline
  && Set (a) && Set (b) & \\
  \hline
 & Edges & FGGM & FGGM-CovSel & FGGM & FGGM-CovSel \\ 
  \hline
1 & (1, 2) & 787.67 (23.03) & 778.05 (22.35) & 1863.79 (36.12) & 1846.51 (35.38) \\ 
  2 & (1, 3) & 766.83 (12.59) & 756.66 (11.19) & 1832.1 (30.07) & 1806.42 (32.86) \\ 
  3 & (2, 3) &  796.32 (18.35) & 784.23 (22.14) & 1799.54 (31.76) & 1785.05 (34.07) \\ 
  4 & (2, 4) &  730.71 (31.39) & 722.14 (30.57) & 1840.36 (41.73) & 1833.68 (40.06) \\ 
  5 & (3, 4) & 730.91 (31.67) & 722.5 (30.02) & 1764.23 (35.58) & 1736.83 (38.57) \\ 
  6 & (4, 5) &  719.63 (22.99) & 704.54 (20.41) & 1974.63 (34.74) & 1949.14 (32.47) \\ 
  7 & (4, 6) & 738.21 (31.77) & 721.55 (33.27) & 1789.22 (36.9) & 1776.31 (36.65) \\ 
  8 & (5, 6) & 752.9 (32.11) & 742.87 (28.27) & 1793.26 (29.31) & 1781.21 (29.65) \\ 
  9 & (6, 7) & 781.46 (14.92) & 759.19 (20.57) & 1293.24 (32.4) & 1283.55 (31.91) \\ 
  10 & (6, 8) & 778.68 (19.86) & 747.99 (23.62) & 1409.48 (31.34) & 1394.82 (30.09) \\ 
  11 & (7, 8) &  786.81 (21.97) & 751.28 (22.99) & 1910.51 (28.87) & 1880.7 (30.58) \\ 
  12 & (8, 9) & 771.1 (26.52) & 751.06 (32.81) & 2004.36 (24.73) & 1982.75 (29.94) \\ 
  13 & (9, 10) &  771.71 (23.98) & 749.55 (24.65) & 1548.35 (36.78) & 1541.75 (36.25) \\
   \hline
\end{tabular}}
\caption{\label{tab:kl}\black{The KL divergence -mean (sd) across $25$ different replicates between true and estimated cross-covariance matrices (through FGGM and FGGM-CovSel) corresponding to all edges for Simulation sets (a) and (b)}}
\end{table}

\section{Analysis of functional neuroimaging data}
We illustrate the utility of the approach on multivariate functional Magnetic Resonance Imaging (fMRI) data from a study of thermal pain. The data is at the brain-region level (each variable corresponds to a region), and a graph between the variables represents the connectivity between different brain regions. In this study, the left forearm of 33 healthy right-handed participants was exposed to various levels (temperatures) of noxious heat stimuli. Participants gave their informed consent, and the Columbia University Institutional Review Board approved the study. Each participant completed seven runs within a session, consisting of 58-75 trials. During each trial, thermal stimuli were applied to the inner surface of the left forearm. Each stimulus lasted 12.5 seconds, including a three-second ramp-up, two-second ramp-down, and 7.5 seconds at the target temperature.

We collected a total of 1845 functional images (TR$=2000 \, ms$) from each participant using a 3T Philips Achieva TX scanner at Columbia University. Structural images were obtained using high-resolution T1 spoiled gradient recall (SPGR) imaging, while functional echo planar images (EPIs) were acquired with specific parameters; see \cite{woo2015distinct} for details. The structural images were aligned with the mean functional image using an iterative mutual information-based algorithm in SPM8 and then normalized to Montreal Neurological Institute (MNI) space using SPM8's generative segment-and-normalize algorithm. Before preprocessing the functional images, the first four volumes were discarded to ensure image intensity stabilization. Outliers were identified using the Mahalanobis distance, and slice-timing differences were corrected. Motion correction was performed using SPM8, and images were transformed to SPM's normative atlas using warping parameters estimated from the aligned structural images. Finally, the functional images were smoothed with an 8 mm full width at half maximum (FWHM) Gaussian kernel and subjected to a high-pass filter with a cutoff of 180s. For a more comprehensive description of the data acquisition and preprocessing methods, please refer to the work by \cite{woo2015distinct}.

We used a variant of the Yeo atlas \citep{yeo2011organization} to first subdivide the brain into 286 separate non-overlapping brain regions, and thereafter organized these regions into 18 functional networks (17 cortical and one subcortical).  These were further arranged into Visual (VIS), Motor (MOT), Dorsal Attention (DAN), Ventral Attention (VAN), Frontoparietal (FP), Limbic (LIM), Default Mode (DMN), and Subcortical (SC) networks.  Our framework models this dataset as a $286$-dimensional functional data (corresponding to the $q=286$ brain regions) for $n=33$ subjects. For computational efficiency, we reduce the number of time points from $1845$ to $184$ by binning observation within a $10$ time unit interval. 

Our goal is to use this dataset for a proof-of-concept analysis 
to judge the covariance estimation performance of our method FGGM-Covsel, which uses the knowledge of a preestimated connectivity graph among the $286$ regions, compared to the FGGM method, which will re-estimate the graph while estimating the covariance function. \black{In previous fMRI studies, partial separability has been both assumed and confirmed \citep{zapata2019partial}. In a real-world application, one can assess this assumption by verifying whether the correlations among basis coefficients at different levels for different processes are close to zero. Given that our analysis serves only as a proof-of-concept, that there is past evidence supporting partial separability in fMRI data, and that our simulations demonstrated robustness of our approach to violation of the partial separability assumption, we opt not to scrutinize this assumption in this paper to maintain brevity in the discussion.}
We use a random subset of the data to estimate the connectivity graph. 
We randomly split the group of subjects into two sets of roughly equal size - (i) train ($17$ subjects) and (ii) test ($16$ subjects). On the training set, we use FGGM of \cite{zapata2019partial} to learn the graph between brain regions and estimate the covariance function. Then on the test set, we use both FGGM and FGGM-Covsel to estimate covariance. The FGGM reestimates a (possibly) different graph on the test set, while FGGM-CovSel conducts graph-constrained analysis using the graph estimated from the training set. We then compare the covariance estimate of FGGM and FGGM-CovSel in the test set with the FGGM estimate obtained in the train set. We perform this random split $10$ times to obtain cross-validated results. 

\begin{figure}[h]
\begin{center}
\includegraphics[scale=1.2,trim={0 0 0 0cm},clip]{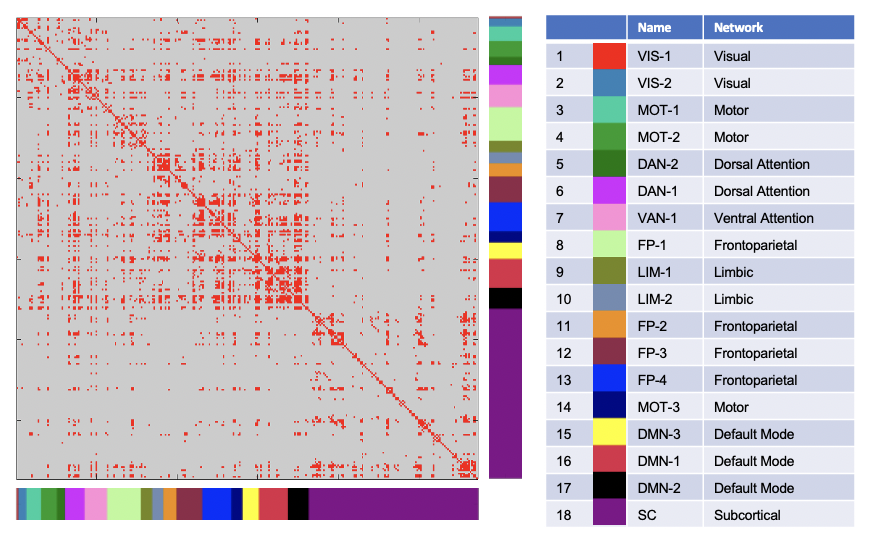}
\end{center}
 \caption{ The final adjacency matrix, consisting of the pairs with an average edge probability larger than $0.5$. The brain is subdivided into $286$ regions using a variant of the Yeo atlas \citep{yeo2011organization}. These regions are organized into 18 networks.  These can be further arranged into Visual (VIS), Motor (MOT), Dorsal Attention (DAN), Ventral Attention (VAN), Frontoparietal (FP), Limbic (LIM), Default Mode (DMN), and Subcortical (SC) networks. The legend to the right illustrates the colors associated with each subnetwork}
    \label{fig:data-adjacency}
\end{figure}

Figure \ref{fig:data-adjacency} visualizes the network between brain regions from the most probable graph across $10$ splits. We average the adjacency matrix across the different data splits and keep the edges in the final graph, as the pairs which have an average edge probability bigger than $0.5$. Pain is known to engage multiple brain networks. 
\textcolor{black}{In our estimated brain network, we found connections among regions within VIS, DAN, VAN, FP, LIM, and DMN.  For example, the DAN may become engaged when individuals are asked to direct attention to pain-related stimuli or when they are actively engaged in pain processing tasks, and the FP network processes subjective and emotional aspects of pain and pain perception \citep{ong2019role}, the DMN modulates pain perception \citep{baliki2014functional}, and the limbic network mediates self-regulation of pain \citep{woo2015distinct}. In addition, subcortical regions regulate pain perception and processing through autonomic responses \citep{bingel2002subcortical} and the somatomotor network is involved in the perception and processing of pain-related sensory input \citep{kropf2018anatomy}.  In addition, connections were found between DMN, FP, LIM, DAN, and VAN, with some engagement with MOT.  Particularly, we see connections between the DMN, DAN, and the FP network. In the context of pain tasks, both these relationships depend on factors such as task demands, attentional focus, and individual differences in pain processing and coping strategies. In general, these networks and connections are consistently activated by experimental pain, illustrating the method's efficacy.}

\begin{figure}[h]
\begin{center}
\includegraphics[scale=1,trim={0 0 0 0cm},clip]{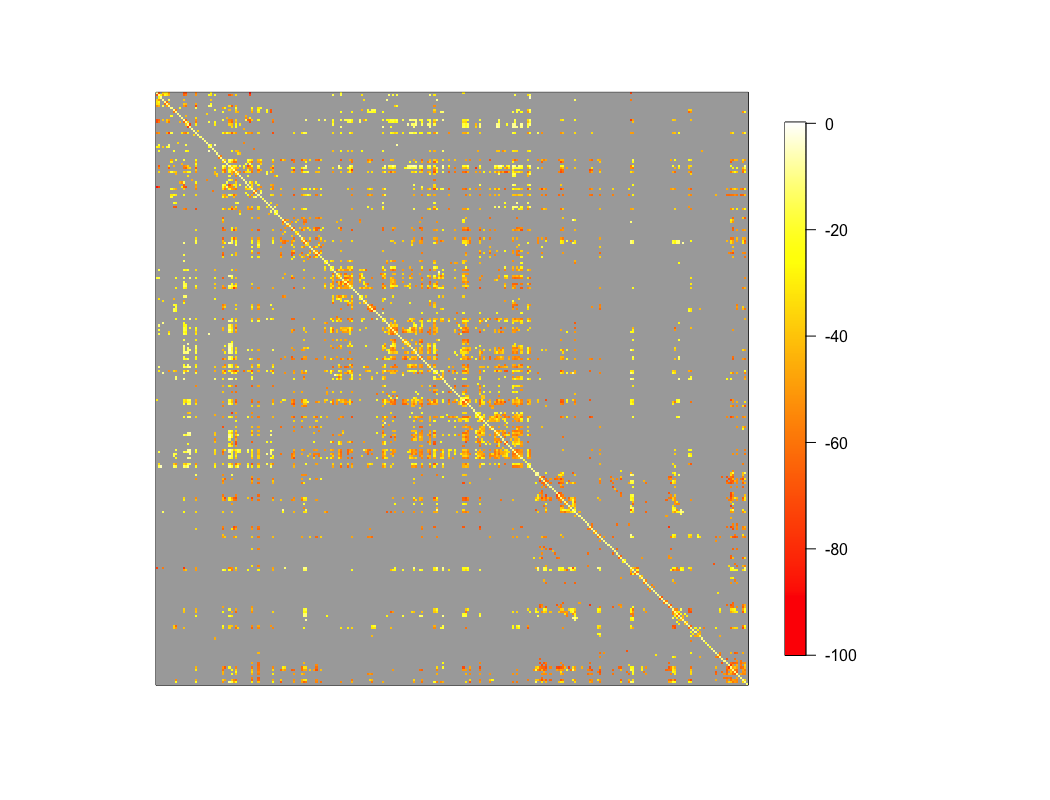}
\end{center}
 \caption{The difference in KL-distances between edge-specific covariances for FGGM-CovSel (test) \& FGGM (train) and FGGM (test) \& FGGM (test). The difference is aggregated across $10$ replicates whenever that specific edge-pair is detected. Grey cells indicate pairs that didn't appear in any edge in $10$ replicates.}
    \label{fig:data-kl-diff}
\end{figure}

We then assess how our graph-constrained analysis (FGGM-Covsel) using this graph compares with FGGM in terms of estimating the covariance function. Let $\widehat C_{train}$ denote the estimated covariance matrix on the training dataset, and $\widehat C_{train}(i,j)$ denote the sub-matrix corresponding to the nodes $i$ and $j$. Similarly, we denote by $\widehat C_{test,m}$ the estimated covariance matrix on the test set by method $m$, $m \in \{\mbox{FGGM,FGGM-CovSel}\}$ and define $\widehat C(i,j)_{test,m}$ accordingly. We then consider the KL distance $d_{KL}(\widehat C(i,j)_{train},\widehat C(i,j)_{test,m})$ for all edges $(i,j)$ in the graph. A lower KL distance implies a better estimation of the covariance. 
Figure \ref{fig:data-kl-diff} plots a heatmap of the difference $$d_{KL}(\hat C(i,j)_{train},\hat C(i,j)_{test,FGGM\mbox{-}CovSel}) - d_{KL}(\hat C(i,j)_{train},\hat C(i,j)_{test,FGGM}).$$ 
The result is aggregated across the replicates in which a specific pair has been detected as an edge by FGGM (train). We observe in Figure \ref{fig:data-kl-diff} that all the values of this difference are negative, implying that the covariance estimate from FGGM-Covsel (test) has a lower KL distance with FGGM (train) than FGGM (test) has with FGGM (train). This is true across all edge-pairs that were included in the connectivity graph. Across $10$ different seeds, the KL distance between FGGM-Covsel (test set) and FGGM (train) is $36.8 (35, 39.9)$ units lower than the distance between FGGM (test) and FGGM (train) when averaged across all edge pairs. These findings show us that estimating the covariance function benefits from the knowledge of the graph and demonstrates the utility of a graph-constrained multivariate functional analysis.

\section{Discussion}

The current manuscript has developed graph-constrained modeling and analysis for multivariate functional data. We build upon recent developments in graphical Gaussian processes (GGP) in the context of highly multivariate spatial data analysis and draw some insightful connections of such frameworks with functional Gaussian graphical models. Our key result is an optimal partially separable functional Gaussian graphical model that is derived using the graphical Gaussian process and preserves the conditional dependencies, or lack thereof, posited by a given undirected graph. Our methodological developments are accompanied by novel theoretical insights connecting the two seemingly disparate paradigms of functional Gaussian graphical models and Graphical Gaussian processes. These theoretical developments enable us to construct a new algorithm to evaluate the maximum likelihood estimate for a partially separable graphical Gaussian model while preserving the conditional dependence constraints imposed by the given graph. This algorithm constitutes the generalization of Dempster's covariance selection from vector-valued data to functional data. \black{Beyond the population level theoretical connections used to derive the methodology, we also prove the consistency of the covariance operator estimator obtained from the resulting algorithm. We acknowledge that this consistency result is only established in a simplistic setting with the curves (functional data) observed at a fixed set of points and the basis functions defined on this discrete domain. Future work needs to refine the theoretical results by filling the remaining gaps between the current theory and the practice. These would need to consider basis functions on a continuous domain and the curves being sampled at a finite set of points in this domain in an increasingly dense manner, i.e.,
in-fill asymptotics. The use of data-driven basis functions from a functional PCA must also be accommodated. Besides consistency, asymptotic normality results would also be needed to conduct inference on the estimated covariance functions.}

We note that while the manuscript focuses on the estimation of the multivariate covariance function under graphical constraints, assuming the mean of the functional processes to be zero, a non-zero mean can be easily accommodated in the framework. This will be suitable in applications where other covariates are observed in addition to the functional responses, and a mean, specified as a linear or non-linear function of these covariates, can be easily estimated as part of the algorithm. \black{Also, our paper focuses on the analysis of multivariate functional data when the within-variable graph is known with certainty. In examples, where the graph is known with some degree of  uncertainty, a Bayesian approach would be more suitable, 
incorporating information about graph uncertainty as a prior for the graphical model. We would then need to adapt our approach to Bayesian reversible jump MCMC algorithms \citep{green2013sampling, barker2013bayesian} similar to what has been done in Section 4.3 of \cite{dey2021ggp}. This will be an important future direction.} 

\black{Our data analysis was conducted under the assumption of partial separability, which necessitates a discussion regarding the practical validity of this assumption. This assumption was motivated by the analysis of similar fMRI data in \cite{zapata2019partial}. However, due to our limited sample size, we cannot explicitly test this assumption in our dataset, as \cite{zapata2019partial} performed in their paper. Hence, we have undertaken an empirical assessment to evaluate the robustness of our methodology under potential violations of the partial separability assumption (Simulation set (b)). This assessment is of utmost importance as it clearly explains our approach's performance in adverse scenarios. While performance indeed declines in scenarios where the assumption is not met, our approach still outperforms the FGGM methods, which also rely on partial separability. It is important to note that when the partial separability assumption is severely violated, FGGM methods such as those proposed by \cite{zhu2016bayesian} may be employed. However, these methods require joint estimation of the entire precision matrix of all regression coefficients, which can lead to significant computational challenges.
}

Given the continued interest in high-dimensional statistical inference for multivariate functional data, this work is expected to spur further research in scaling up the proposed models here to massive numbers of functional inputs. Specific examples include settings where spatial data are measured on a very large number of dependent outcomes over massive numbers of locations. In this regard, it will be possible to embed Gaussian processes \citep[see, e.g.,][for hierarchical Nearest Neighbor Gaussian process models]{datta2016hierarchical,datta2016nearest,finley2019efficient} within the nodes of the graph to scale up to the massive numbers of locations. 

While the conceptual idea of building scalable multivariate models with stochastic processes as nodes of a graph follows from graphical Gaussian processes \cite{dey2021ggp} or from approaches for versatile classes of multivariate Markov random fields \citep{jin2007order}, substantial developments on precise model specifications and, perhaps even more importantly, algorithmic developments are still warranted. In this regard, we expect our current developments to generate future research pursuits that will culminate in a versatile, computationally efficient framework for graph-constrained analysis of high-dimensional dependent functional or spatial data. Applications of such a framework will span biomedical and health sciences, broader environmental sciences, social sciences, and economics. 

\section*{Acknowledgement}
AD was partially supported by the National Science Foundation (NSF) Division of Mathematical Sciences grant DMS-1915803 and National Institute of Environmental Health Sciences (NIEHS) grant R01 ES033739. SB was supported by the NSF awards DMS-1916349 and DMS-2113778 and by the National Institute of General Medical Sciences (NIH-NIGMS) awards R01GM148761-01. DD undertook this project during his PhD at Johns Hopkins Bloomberg School of Public Health and was supported by a fellowship through a Joint Graduate Training Program, established in collaboration between the Department of Biostatistics at the Johns Hopkins Bloomberg School of Public Health and the Intramural Research Program of the National Institute of Mental Health. The authors thank Dr. Brian Caffo for helpful discussions. 

\appendix

\section{Proof of Theorem \ref{thm:ggp-fggm-infinite}}
\label{appn:proof-thm1}
\begin{proof}
Denote $\Sigma_l = (\sigma_{lij})$, $\Sigma^*_l = (\sigma^*_{lij})$, and $\Sigma_l ^{*-1} = \Omega_l = (\omega^*_{lij})$. The cross-covariance function of $w(\cdot)$ is denoted by $C^*=(C^*_{ij})$ and the conditional cross-covariance function is $K^*=(K_{ij}^*)$. By Theorem~3 of \cite{zapata2019partial}, 
\begin{equation}\label{eq:condcross}
    K^*_{ij}(s,t) = \sum_{l=1} ^{\infty} \frac{-\omega^*_{lij}}{\omega^*_{lii}\omega^*_{ljj} - \omega_{lij}^{*2}} \phi_l(s) \phi_l(t).
\end{equation}
\noindent (If part:) Suppose $ \Sigma^*_l=\mbox{CovSel}(\Sigma_l,\calG)$ for all $l$.  To prove $w(\cdot) \sim GGP(C,\calG)$ we first show it is a graphical GP. For any $(i,j) \notin E$, by the property of covariance selection on each $\Sigma_l$, we have $\omega^*_{lij}=0$ for all $l$. Hence, from (\ref{eq:condcross}) $K^*_{ij}(s,t)=0$ for all $s,t \in \calD$, $(i,j) \notin E$ and $w(\cdot)$ is a graphical GP with respect to the graph $\calG$. Next we show $w(\cdot)$ retains the marginals and edge-specific cross-covariances from $C$. This is immediate as once again for all $(i,j) \ni i=j$ or $(i,j) \in E$, we have by the property of covariance selection
$\sigma^*_{lij}=\sigma_{lij}$ and consequently $C^*_{ij}(s,t) = \sum_{l=1}^\infty \sigma^*_{lij}\phi_l(s)\phi_l(t) = \sum_{l=1}^\infty \sigma_{lij}\phi_l(s)\phi_l(t) = C_{ij}(s,t)$. \\

\noindent (Only if part:) Let $w(s) = \sum_{l=1}^\infty \theta^*_l \phi_l(s) \sim GGP(C,\calG)$. Since it is a graphical GP, then $K_{ij}^*(s,t)=0$ for all $s,t \in \calD$ and $(i,j) \notin E$. By orthogonality of $\phi_l$'s, $$ 0 = \int K_{ij}^*(s,t)\phi_l(s)ds = \frac{-\omega^*_{lij}}{\omega^*_{lii}\omega^*_{ljj} - \omega_{lij}^{*2}} \phi_l(t).$$
This holds for all $t \in \calD$ implying $\omega^*_{lij}=0$ for all $l$. By definition (\ref{def:ggp}), $w(\cdot)$ retains the marginal and edge-specific cross-covariances. Hence, for $i=j$ or $(i,j) \in E$, $C^*_{ij}=C_{ij}$ implying 
$$ \sigma_{lij}\phi_l(t) = \int C_{ij}(s,t)\phi_l(s)ds = \int C_{ij}^*(s,t)\phi_l(s)ds = \sigma^*_{lij}\phi_l(t).$$ This holds for all $t$ implying $\sigma_{lij}=\sigma^*_{lij}$ for all $l$, $i=j$ or $((i,j) \in E$. Hence, $\Sigma^*_l$ satisfies all the conditions of covariance selection (Definition \ref{def:covsel}) and is the unique matrix $\mbox{CovSel}(\Sigma_l,\calG)$. \end{proof}

\section{\black{Proof of Corollary 1.3}}\label{sec:cor-proof}

\black{We apply Theorem~3 of \cite{zapata2019partial} to the process $w^m(s)$ and write the conditional cross-covariance function in terms of the elements of $\Sigma_l ^{*-1} = \Omega_l = (\omega^*_{lij})$ and the basis functions as}
\black{
\begin{align*}
    K^m_{ij}(s,t) & = \text{Cov}(w^m_i(s), w^m_j(t) \mid w^{m}_{-(i,j)}) = \text{Cov}\left(\left.\sum_{l=1}^m \theta_{li}^* \phi_l(s), \sum_{l'=1}^m \theta_{l'j}^* \phi_{l'}(t) \right| w^m_{-(i,j)}\right) \\ 
    &= \sum_{l,l'=1}^{m}\text{Cov}\left(\theta_{li}^*, \theta_{l'j}^*  \mid w^m_{-(i,j)}\right) \phi_l(s)\phi_{l'}(t) = \sum_{l=1}^{m}\text{Cov}\left(\theta_{li}^*, \theta_{lj}^*  \mid \theta_{l, -(i,j)}^*\right) \phi_l(s)\phi_{l}(t) \\
    & = \sum_{l=1} ^{m} \frac{-\omega^*_{lij}}{\omega^*_{lii}\omega^*_{ljj} - \omega_{lij}^{*2}} \phi_l(s) \phi_l(t),
\end{align*}}
\black{where $w^{m}_{-(i,j)}$ is the full (uncountable) realization of all the component processes in $w^{(m)}(s)$ except $i$ and $j$. The third equality follows from the fact that the coefficients $\theta_{l}$ and the process $w^{(m)}(s)$ uniquely identify each other given the $m$ basis functions. The proof now follows that of Theorem~\ref{thm:ggp-fggm-infinite}.
}

\section{Approximate equivalence of stitching and low-rank truncation}\label{sec:lowrank}
To derive this approximate result, we observe that  $w_{stitch}^{pp}(s) = D(s)w_{stitch}(\calL)$ where $D(s)=diag(C_{11}(s,\calL) C_{11}(\calL,\calL)^{-1}, \ldots, C_{qq}(s,\calL) C_{qq}(\calL,\calL)^{-1})$ and $w_{stitch}(\calL)$ is a random variable whose distribution is specified in (\ref{eq:covsel}). Next, as $w(\cdot) \sim GGP(C,\calG)$, the component processes $w_j(\cdot) \sim GP(0,C_{jj})$ retain the marginals. When the reference set $\calL$ is chosen to have $m$ locations, the truncated process $w^m_j(\cdot)$ is related to the untruncated process $w_j(\cdot)$ via the following approximation \citep[see, e.g., p. 389-390 in ][ for a derivation ]{banerjee2014hierarchical}
\begin{equation}\label{eq:nystrom}
    w_j^{m}(s) \approx C_{jj}(s,\calL) C_{jj}(\calL,\calL)^{-1}w_{j}(\calL)
 \mbox{ for all } s \in \calD, j \in \calV.
 \end{equation} 
The approximate equivalence in (\ref{eq:nystrom}) is related to the N\"ystrom approximation for kernel matrices \citep{drineas2005nystrom}. Rewriting (\ref{eq:nystrom}) as  $w^m(s) \approx D(s)w(\calL)$ and since $w_{stitch}^{pp}(s) = D(s)w_{stitch}(\calL)$,  to show $w^m(\cdot) \approx w_{stitch}^{pp}(\cdot)$, we now only need  $w(\calL)$ and $w_{stitch}(\calL)$ to  approximately have the same distribution. As $w(\cdot) \sim GGP(C,\calG)$, for $i=j$ or $(i,j) \in E$, we have $\mbox{Cov}(w_i(\calL),w_j(\calL)) =  C_{ij}(\calL,\calL) = \mbox{Cov}(w_{stitch,i}(\calL),w_{stitch,j}(\calL)) $ by definition of covariance selection. Finally, we show that for $(i,j) \notin E$, $ \mbox{Cov}(w_i(\calL),w_j(\calL) \mid w_{-(ij)}(\calL)) \approx O = \mbox{Cov}(w_{stitch,i}(\calL),w_{stitch,j}(\calL) \mid w_{stitch,-ij}(\calL))$ as follows:
\begin{align*}
& Cov(w_i(\calL), w_j(\calL) \mid w_{-(ij)}(\calL)) 
= Cov\left(C_{ii}(\calL,\calL) C_{ii}(\calL,\calL)^{-1}w_{i}(\calL),\right.\\
& \qquad \left. C_{jj}(\calL,\calL) C_{jj}(\calL,\calL)^{-1}w_{j}(\calL) \mid C_{kk}(\calL,\calL) C_{kk}(\calL,\calL)^{-1}w_{k}(\calL), k \in \calV \setminus \{i,j\}\right)\\ 
&\quad \approx Cov(w^m_i(\calL), w^m_j(\calL) \mid w_{-ij}(\calL)) \mbox{ by (\ref{eq:nystrom})} \\
&\quad = Cov(w^m_i(\calL), w^m_j(\calL) \mid C_{kk}(s,\calL) C_{kk}(\calL,\calL)^{-1}w_{k}(\calL), s \in \calD, k \neq i,j)\\
& \quad \approx  Cov(w^m_i(\calL), w^m_j(\calL) \mid w^m_{-ij}(\cdot)) \mbox{ again by (\ref{eq:nystrom})} \\
&\quad = O \mbox{ for } (i,j) \notin E.
\end{align*}
The middle equality follows from the fact that for any $s \in 
\calL$ and any $k$, we have $C_{kk}(s,\calL) C_{kk}(\calL,\calL)^{-1}w_{k}(\calL)=w_k(s)$. Hence, the 
$\sigma$-algebras $\sigma(w_{-ij}(\calL))$ and $\sigma(C_{kk}(s,\calL) C_{kk}(\calL,\calL)^{-1}w_{k}(\calL), s \in \calD, k \neq i,j))$ becomes the same since the latter is generated by the former and contains the former. Corollary~\ref{cor:trunc} ensures the last equality since $w^m(\cdot)$ is a graphical GP conforming to $\calG$. 

We have now shown that the covariances of $w(\calL)$ and $w_{stitch}(\calL)$ agree exactly on the entries corresponding to $i=j$ or $(i,j) \in E$, and the inverse-covariances agree approximately on the entries  corresponding to $(i,j) \notin E$. By uniqueness of Dempster's covariance selection for positive definite matrices, $w(\calL)$ and $w_{stitch}(\calL)$ have approximately the same distribution. 
This establishes the equivalence of the  $m$-truncated partially separable FGGM  $w^m(s)=D(s)w(\calL)$ with a stitched  predictive process $w^{pp}_{stitch}(s)=D(s)w_{stitch}(\calL)$ on $m$ reference locations.

\section{Proof of Theorem \ref{thm:mle}}
\label{appn:proof-thm3}
\begin{proof}
We have $X_i \overset{\mathrm{iid}}{\sim} N_{pq} (0, \black{C^m})$ for $i=1, \cdots, N$, where $\black{C^m} = \sum_{l=1}^{m} \Sigma_l \otimes \boldsymbol{\phi}_l \boldsymbol{\phi}_l '$ is the true multivariate covariance function of the partially separable process. 
We use $X_{ij}$ to denote the $p$-variate vector observed for $i$-th replicate and $j$-th variable. Then, the sample covariance matrix $S_{(pq \times pq)}$ can be calculated as $S = \frac{1}{N}\sum_{i=1}^{N} \begin{pmatrix}
           X_{i1}', 
           X_{i2}', \hdots, X_{iq}'
           \end{pmatrix}'\begin{pmatrix}
           X_{i1}', 
           X_{i2}', \hdots, X_{iq}'
           \end{pmatrix} =\frac{1}{N}\sum_{i=1}^{N} X_iX_i'$. 
Under the constraint of a given graphical model $\calG$, we want to find the maximum likelihood estimate (MLE) of the parameters $\Sigma_l$'s of the partially separable process. 

First we will derive the MLE of a multivariate partially separable covariance function without the graphical constraint.  
Under partial separability, we can write $X_{i} = \sum_{l=1}^m \theta_{il} \otimes \bphi_l$, where $\theta_{il} \sim N(0,\Sigma_l)$. Using orthonormality of the basis functions, we can transform the data as follows $ (I_q \otimes \bphi_l') X_i  = \theta_{il}$. Here, the $X_i$'s (jointly over $i$) and the $\theta_{il}$'s (jointly over $i,l$) are a 1-1 and invertible linear transformation of each other. 


It's easy to check that $\theta_{il}$'s are iid across $i$'s and independent across $l$'s. For $i\ne j$, $\textrm{Cov}(\theta_{il} , \theta_{jl}) = E[(I_q \otimes \bphi_l') X_i X_j' (I_q \otimes \bphi_l)] = 0$, since $E[X_i X_j']=0$. Now, for $l \ne k$, $\textrm{Cov}(\theta_{il} , \theta_{ik}) = E[(I_q \otimes \bphi_l') X_i X_i' (I_q \otimes \bphi_k)] = (I_q \otimes \bphi_l') \black{C^m} (I_q \otimes \bphi_k) = (I_q \otimes \bphi_l') (\Sigma_k \otimes \bphi_k) = 0$. Similarly, for $i \ne j, l \ne k$, $\textrm{Cov}(\theta_{il} , \theta_{jk})=0$. 

The derivation above can help us rewrite the likelihood in following way \begin{equation}
\label{eq:lik-pca}
\begin{split}
L(\Sigma_l; l=1, \cdots, m| X_1, \cdots, X_N) & \propto L(\Sigma_l; l=1, \cdots, m| \theta_{il}; i=1 \cdots, N, l=1, \cdots, m)\\
& \propto \Pi_{l=1}^{m} \Pi_{i=1}^{N} L(\Sigma_l|\theta_{il})    
\end{split}.
\end{equation}
This result \eqref{eq:lik-pca} is due to the fact that the Jacobian transformation in the likelihood going from the $X_i$'s to the $\theta_{il}$'s will only depend on known $\bphi_l$'s. So, finding the maximum likelihood estimate (MLE) of $\Sigma_l$ boils down to maximizing the transformed data-likelihood of iid multivariate normal variables $\theta_{il}, i=1, \cdots, N$. Hence, we get the MLE of the parameters as $\black{\tilde{\Sigma}_l} = S_l = \frac{1}{n} \sum_{i=1}^N  {\theta_{il} \theta_{il}'}$. Rewriting $\theta_{il}$ in terms of $X_i$, we have the MLE as $\black{\tilde{\Sigma}_l}=(I_q \otimes \bphi_l') \frac{1}{N}\sum_{i=1}^N (X_i X_i') (I_q \otimes \bphi_l)= (I_q \otimes \bphi_l') S (I_q \otimes \bphi_l)$. Thus the MLE of a multi-variate partially separable covariance function without any graphical constraint is given by
\begin{equation}\label{eq:mle}
    \widehat{C^m}_{unconstrained} = \sum_{l=1}^{m} (I_q \otimes \bphi_l') S (I_q \otimes \bphi_l) \otimes {\bphi}_l  {\bphi}_l'
\end{equation}

We now derive the MLE under the graphical constraint. Based on the results of Theorem \ref{thm:ggp-fggm-infinite} and Corollary \ref{cor:trunc}, a natural guess would be that the MLE under the graphical constraint can be obtained by applying covariance selection on each of the estimates $\black{\tilde{\Sigma}_l= S_l} = (I_q \otimes \bphi_l') S (I_q \otimes \bphi_l)$ in (\ref{eq:mle}). We prove this formally below. 

If we assume the variables are conforming to a graph $\calG = (V, E)$, i.e., $\Sigma_l \in M^+ (\calG) = \{S \in \mathbb R^{q \times q}: S_{ij} = 0\, \forall\, (i,j) \notin E\}$ for all $l=1, \cdots, m$, then by \cite{dempster1972covariance}, we get graph-constrained MLE  by maximizing the likelihood in (\ref{eq:lik-pca}) under the constraint
$\{\Sigma_1, \ldots, \Sigma_m: \Sigma_l \succ 0,  \Sigma_l \in M^+ (\calG) \forall l=1,\ldots,m\}$. 
Due to the product-form (across $l$'s) in (\ref{eq:lik-pca}), this becomes equivalent to finding the MLE for each of $\Sigma_l$'s separately under the graphical constraint. In other words, the MLE is given by 
\begin{align*}
\textrm{argmax}_{\Sigma_l \succ 0,  \Sigma_l \in M^+ (\calG)} \Pi_{i=1}^{N} L(\Sigma_l|\theta_{il}) =  \textrm{argmin}_{\Sigma_l \succ 0,  \Sigma_l \in M^+ (\calG)}  [\textrm{tr}({\Sigma}_l^{-1} S_l) + log(\mid\Sigma_l\mid)] 
\end{align*}
Thus the problem has exactly reduced to Dempster's covariance selection problem, i.e., given iid multivariate realizations $\theta_{il}$ finding the MLE for the covariance matrices $\Sigma_l$ under the graphical constraint. The solution is given by $\hat \Sigma_l = CovSel(S_l, \calG)$ which yields the MLE for $\black{C^m}$ as given in (\ref{eq:mlegraph}). 
\end{proof}

\section{\black{Proof of Lemma \ref{lem:con}}}

\label{appn:proof-lem1}
\begin{proof}
\black{Let's denote, for $N$ samples, the sample covariance matrix as $S_N$ and the basis-level covariance matrix estimate for $\Sigma_l$ as $\tilde \Sigma_{lN} = S_{lN} = (I_q \otimes \bphi_l') S_N (I_q \otimes \bphi_l)$ and the corresponding covariance selection estimator as $\hat \Sigma_{lN} = CovSel(\tilde \Sigma_{lN}, \calG)$. Let's define I-divergence between two matrices $A$ and $B$ as $\calI(B||A) = \frac{1}{2}(tr(A^{-1}B - I) - log(A^{-1}B)) = KL(B||A) - \frac{q}{2}$, where $A$ and $B$ are $q \times q$ positive-definite matrices.} 

\black{Since, the $X_i$'s are iid, the sample covariance $S_N$ is an asymptotically consistent estimator for population covariance $C^m$,. Then $\tilde \Sigma_{lN} = (I_q \otimes \bphi_l') S_N (I_q \otimes \bphi_l)$ converges to $ (I_q \otimes \bphi_l') C^m (I_q \otimes \bphi_l) = \Sigma_l$ in probability as $N \rightarrow \infty$. Now we will show that, $\hat \Sigma_{lN} = CovSel(\tilde \Sigma_{lN}, \calG)$ converges to $\Sigma_l$ in probability as well. }

\black{By Lemma 1 of \cite{speed1986gaussian}, $\hat \Sigma_{lN} = argmin_{A: A \succ 0,  A \in M^+ (\calG)} \calI(A || \tilde \Sigma_{lN})$. Since, $\Sigma_l \succ 0, \Sigma_l \in M^+ (\calG)$, we have $\calI(\hat \Sigma_{lN}|| \tilde \Sigma_{lN}) \leq \calI(\Sigma_l|| \tilde \Sigma_{lN})$. Using the fact that, $\tilde \Sigma_{lN}$ converges to $\Sigma_l$ in probability, we get $\tilde \Sigma_{lN}^{-1} \Sigma_l \rightarrow I_q$ in probability and $I(\Sigma_l|| \tilde \Sigma_{lN}) \rightarrow 0$. Hence, $\calI(\hat \Sigma_{lN}|| \tilde \Sigma_{lN}) \rightarrow 0$. Using Lemma 1, part (iii) of \cite{speed1986gaussian}, we get, $\hat \Sigma_{lN} - \tilde \Sigma_{lN} \rightarrow 0$ in probability. Using Slutsky's theorem, we get $\hat \Sigma_{lN} = CovSel(\tilde \Sigma_{lN}, \calG) = CovSel(S_{lN}, \calG)$ converges to $\Sigma_l$ in probability.}

\black{Using Slutsky's theorem again, we get $\black{\widehat{C^m}} = \sum_{l=1}^{m} \hat \Sigma_{lN}  \otimes {\bphi}_l {\bphi}_l'$ converges to $\sum_{l=1}^{m} \Sigma_l \otimes {\bphi}_l {\bphi}_l' = C^m$ in probability.}
\end{proof}

\section{Supplementary Figures}

\begin{figure}[H]
    \begin{center}
 \subfloat[Variable 2 \label{fig:sim-res-stitch-2}]
{\includegraphics[scale=0.35,trim={0 0 0 1cm},clip]{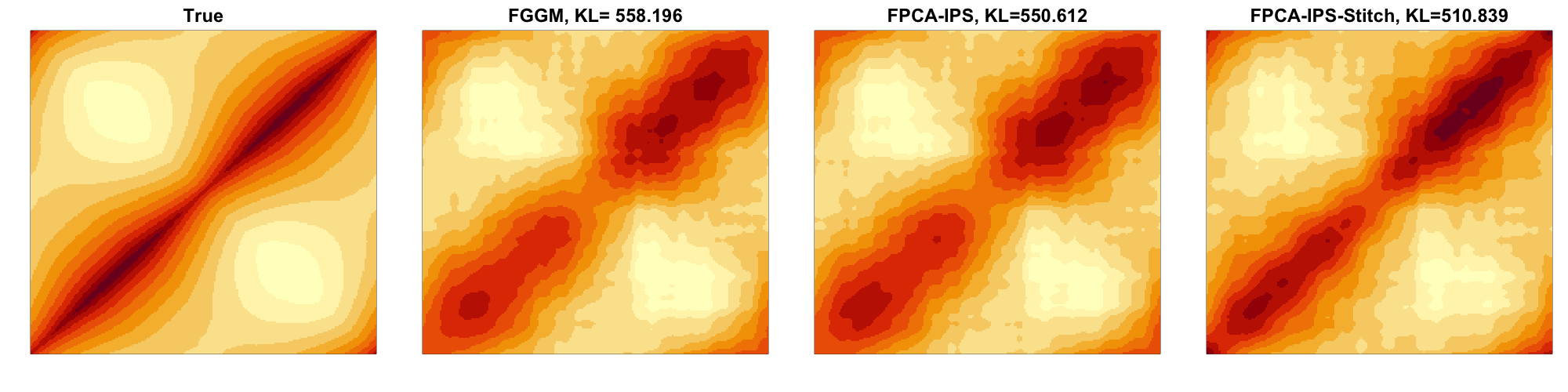}}
\hspace{0.05cm}
\subfloat[Variable 3 \label{fig:sim-res-stitch-3}]
{\includegraphics[scale=0.35,trim={0 0 0 1cm},clip]{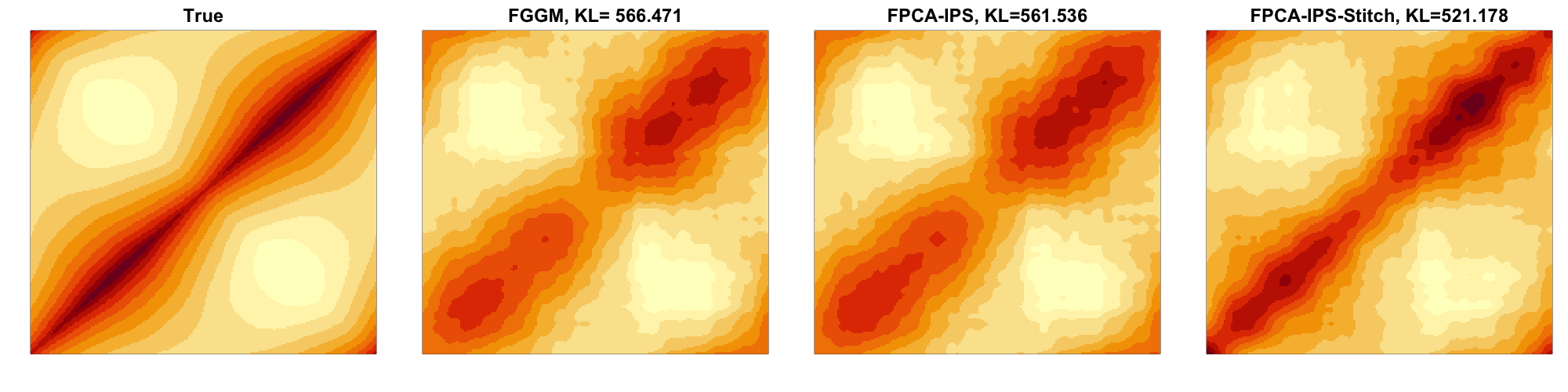}}\\
\subfloat[Variable 4 \label{fig:sim-res-stitch-4}]
{\includegraphics[scale=0.35,trim={0 0 0 1cm},clip]{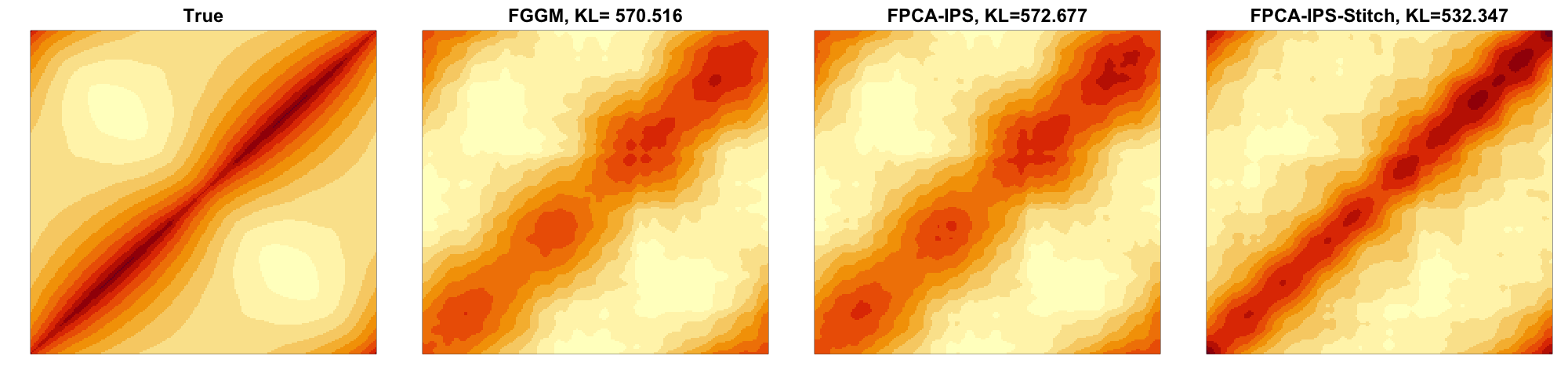}}
\hspace{0.05cm}
\subfloat[Variable 5 \label{fig:sim-res-stitch-5}]
{\includegraphics[scale=0.35,trim={0 0 0 1cm},clip]{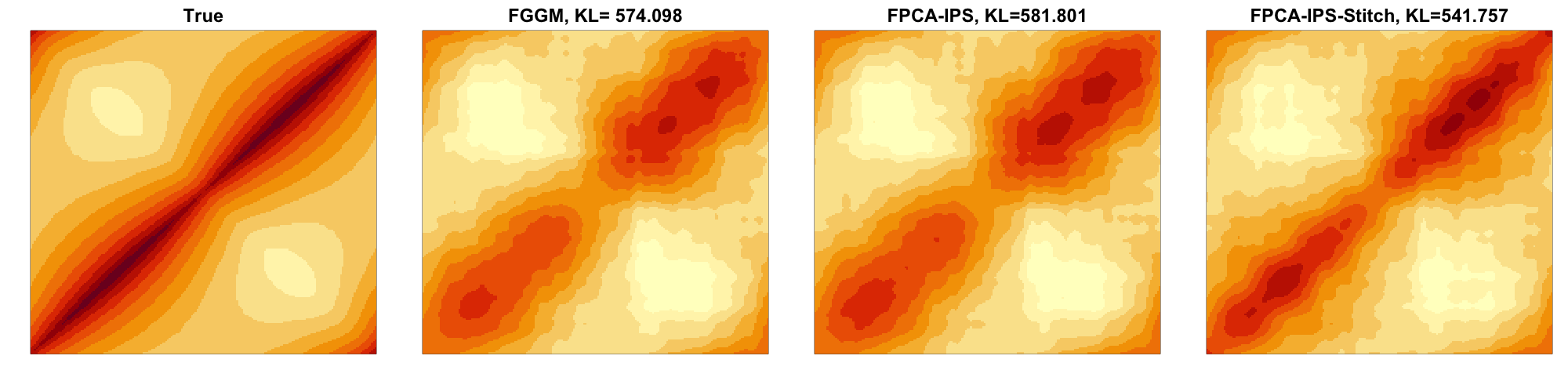}}\\
\subfloat[Variable 6 \label{fig:sim-res-stitch-6}]
{\includegraphics[scale=0.35,trim={0 0 0 1cm},clip]{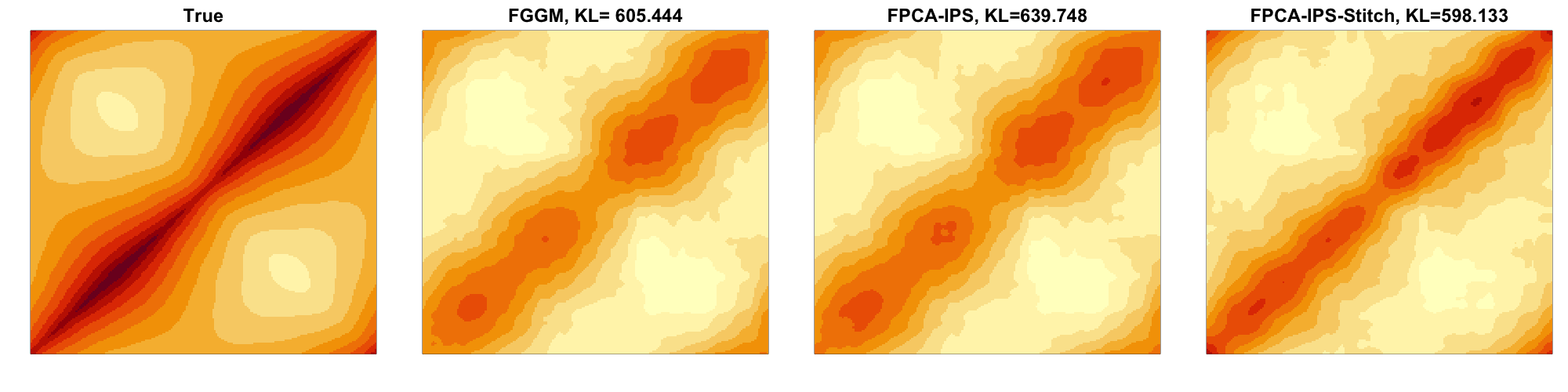}}
\hspace{0.05cm}
\subfloat[Variable 7 \label{fig:sim-res-stitch-7}]
{\includegraphics[scale=0.35,trim={0 0 0 1cm},clip]{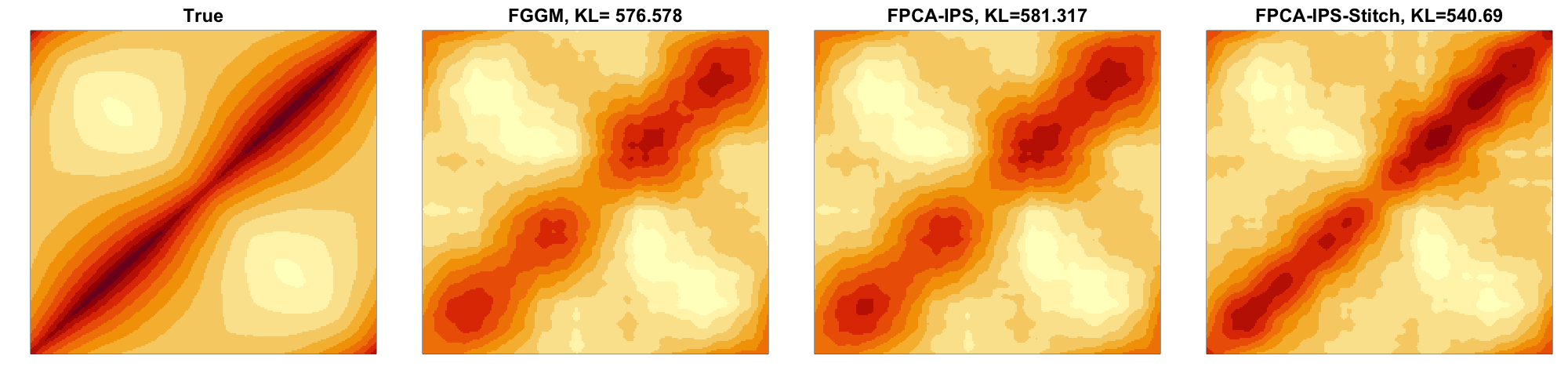}}\\
\subfloat[Variable 8 \label{fig:sim-res-stitch-8}]
{\includegraphics[scale=0.35,trim={0 0 0 1cm},clip]{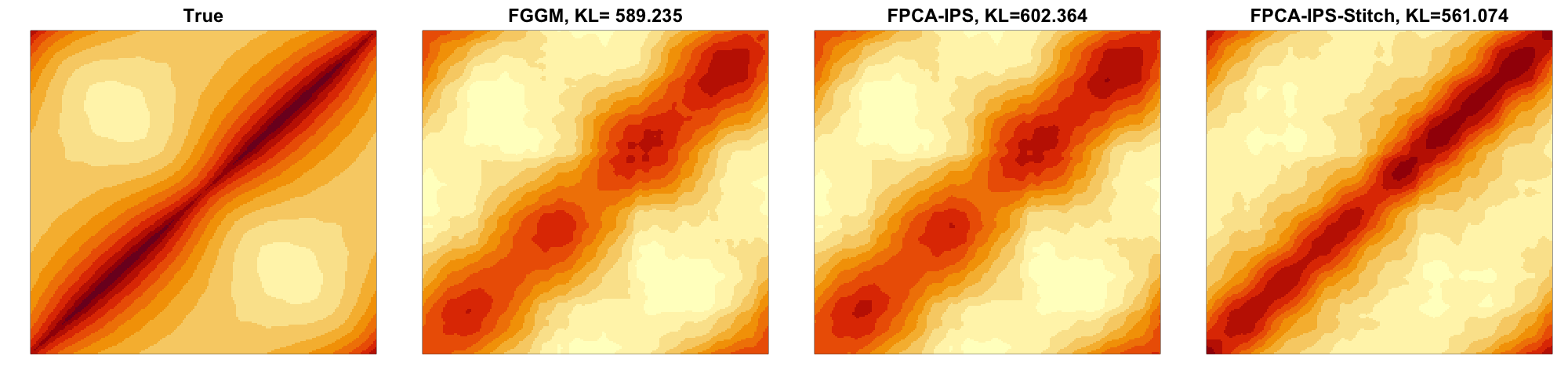}}
\hspace{0.05cm}
\subfloat[Variable 9 \label{fig:sim-res-stitch-9}]
{\includegraphics[scale=0.35,trim={0 0 0 1cm},clip]{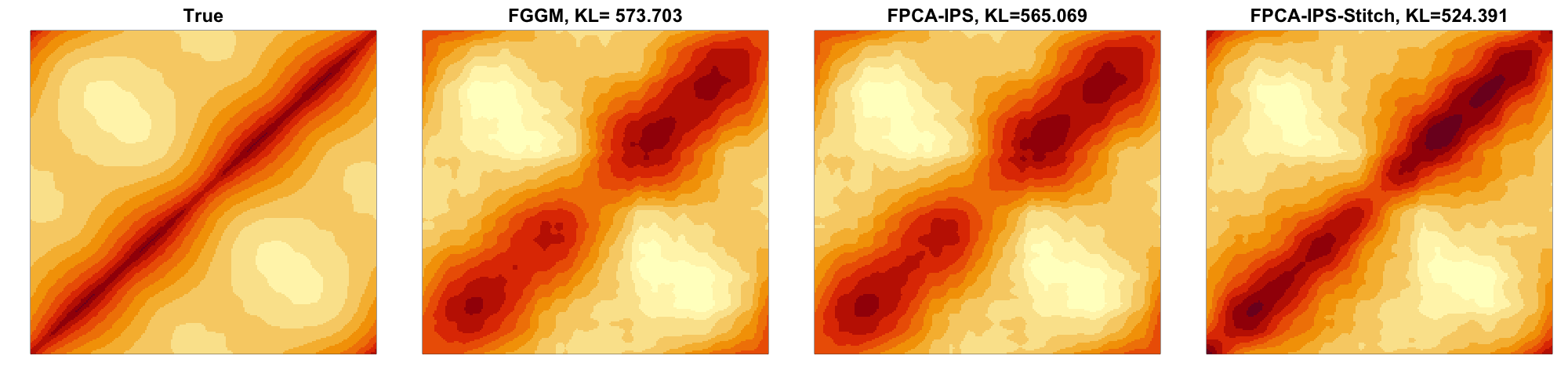}}\\
\subfloat[Variable 10 \label{fig:sim-res-stitch-10}]
{\includegraphics[scale=0.35,trim={0 0 0 1cm},clip]{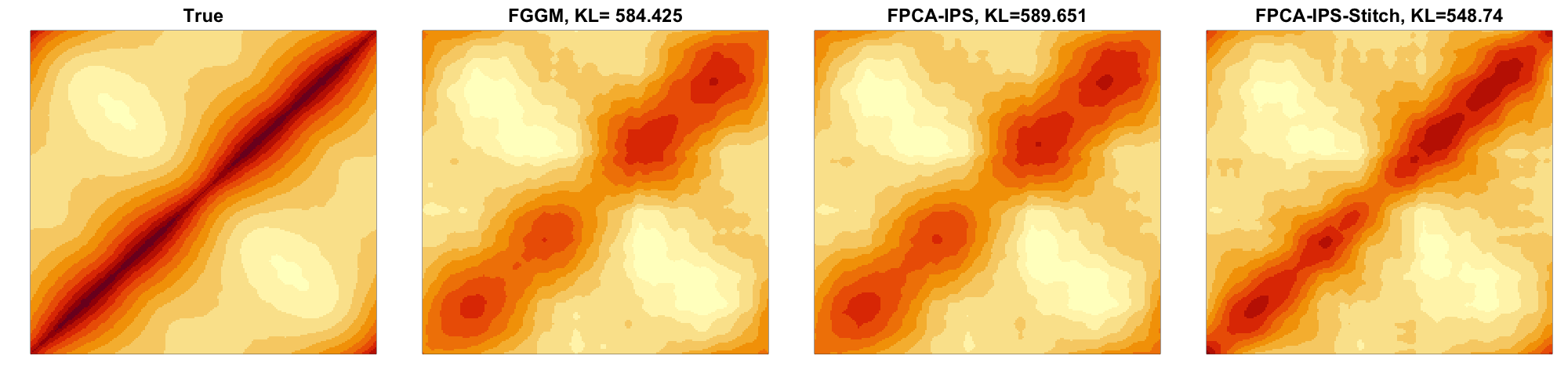}}
\end{center}
 \caption{Marginal covariance surfaces for variables $2 - 10$ in Set (a): True, FGGM, FGGM-CovSel, \black{FGGM-Stretch} (from left to right).}
    \label{fig:sim-res-stitch-rest}
\end{figure}

\begin{figure}[H] 
    \begin{center}
 \subfloat[\black{set (a)} \label{fig:graph-est-a}]
{\includegraphics[scale=0.6,trim={0 0 0 0cm},clip]{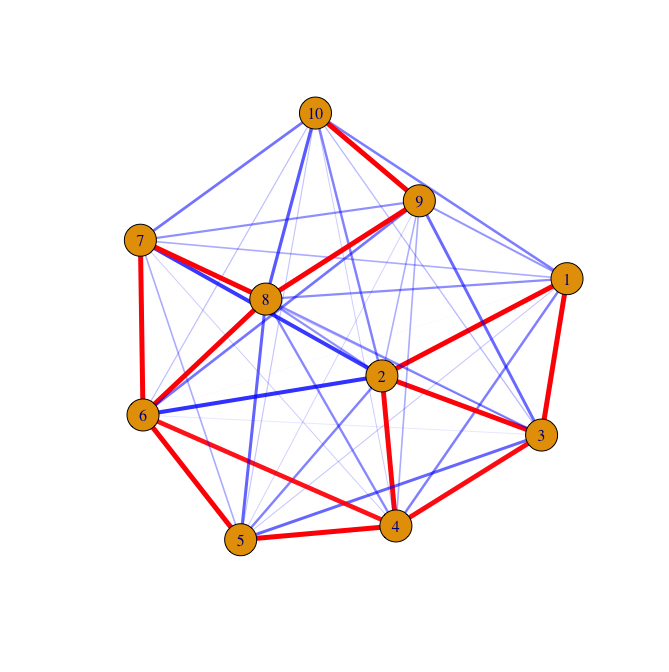}}
\hspace{0.3cm}
\subfloat[\black{set (b)} \label{fig:graph-est-b}]
{\includegraphics[scale=0.6,trim={0 0 0 0cm},clip]{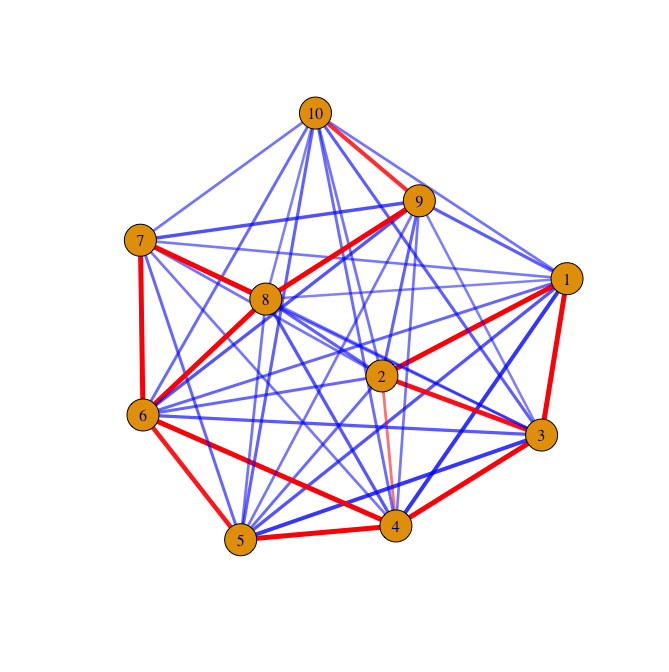}}\\
 \end{center}
 \caption{\black{Most probable graphs estimated through FGGM in set (a) and set (b). Edge widths and opacity are proportional to edge selection probability (averaged over seeds). Edges in the true graph are colored red, otherwise the edges are colored blue.} 
 }
 \label{fig:graph-est}
\end{figure}

\begin{figure}[H]
\begin{center}
\includegraphics[scale=1,trim={0 0 0 0cm},clip]{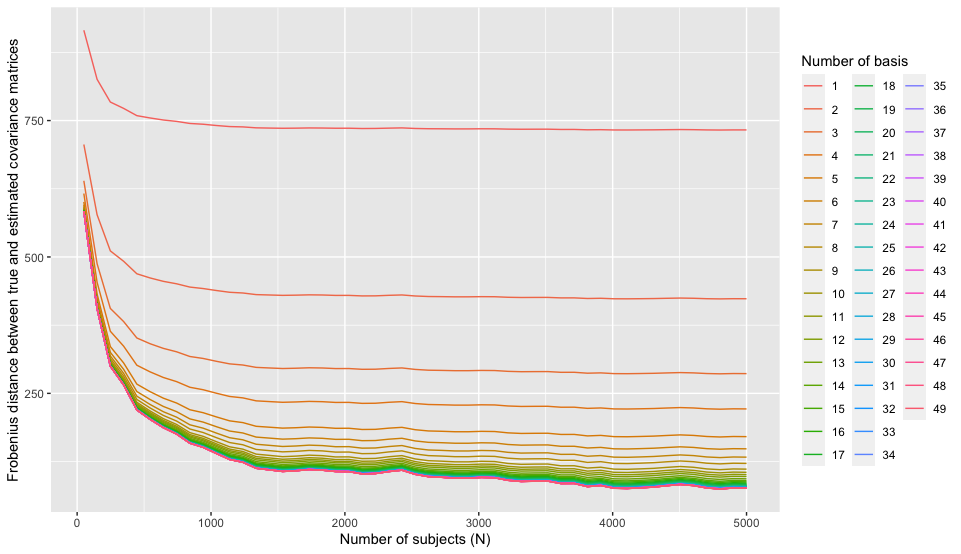}
\end{center}
 \caption{\black{The difference (in Frobenius norm) between true and FGGM-Covsel estimated covariance matrix ($2000 * 2000$) with increasing sample size and number of basis functions}}
    \label{fig:asymp-frob}
\end{figure}

\bibliographystyle{biom} 
\bibliography{GGP_PS_Arxiv_092424}

\end{document}